\documentclass[11pt,a4paper]{article}
\usepackage[utf8]{inputenc}
\usepackage{amsmath}
\usepackage{placeins}
\usepackage{url}
\usepackage{natbib}
\usepackage[table]{xcolor}
\usepackage{setspace}
\usepackage{geometry}
\usepackage{cite}
\usepackage{amsfonts}
\usepackage[normalem]{ulem}
\usepackage{pifont}
\usepackage{hyperref}

\usepackage{eurosym}

\usepackage{tikz}
\usetikzlibrary{shapes}
\usetikzlibrary{plotmarks}
\usetikzlibrary{tikzmark,matrix,calc}
\usetikzlibrary{arrows.meta,calc,decorations.markings,math,arrows.meta}
\usetikzlibrary{decorations.pathreplacing}
\usepackage{booktabs}   
\usepackage{siunitx}    

\usepackage{arydshln}
\usepackage{amssymb}
\usepackage{multirow}
\usepackage{graphicx}
\usepackage{soul}
\soulregister\cite7
\soulregister\ref7
\usepackage{rotating}
\usepackage{threeparttable}
\usepackage{authblk}
\usepackage{subcaption}

\usepackage{appendix}
\usepackage{bm}

\usepackage{amsthm}
\usepackage{pdflscape}
\usepackage{array}
\usepackage{upgreek}
\usepackage{enumitem}
\usepackage{tabularx}
\usepackage{algorithm}
\usepackage{algorithmic}
\usepackage{helvet}
\usepackage{caption}

\usetikzlibrary{positioning}

\newcolumntype{C}{>{\centering\arraybackslash}X}

\newcommand{\ignore}[1]{}

\begin{document}
\begin{spacing}{1.5}	

\title{Integrity at Stake: Statistical Detection of\\ Anomalous In-Play Betting in Football}

\author[$\dag$ $\ddag$ $\S$ $\star$]{David Winkelmann}
\author[$\dag$ $\ddag$]{Maya Vienken}
\author[$\dag$]{Christian Deutscher}
\author[$\ddag$]{Roland Langrock}

\affil[$\dag$]{Department of Psychology and Sports Science, Bielefeld University, Bielefeld, Germany}
\affil[$\ddag$]{Department of Business Administration and Economics, Bielefeld University, Bielefeld, Germany}
\affil[$\S$]{Department of Business Decisions and Analytics, University of Vienna, Vienna, Austria}

\affil[$\star$]{Corresponding author: david.winkelmann@univie.ac.at}

\maketitle
\noindent
\normalsize

\begin{abstract}
Match-fixing undermines the integrity of sport by eroding public trust and threatening the financial sustainability of clubs and leagues. The expansion of sports betting markets has created new incentives and opportunities for manipulation, calling for rigorous, data-driven monitoring tools. Football, which accounts for the largest share of global betting turnover, remains particularly exposed: integrity reports continue to flag several suspicious matches, with past scandals in Italy and Turkey underlining the problem's persistence. This study uses high-frequency in-play (live) betting data from the Italian Serie B (seasons 2018/19--2020/21) to explore statistical approaches for detecting unusual betting behaviour. A state-space modelling framework is employed to describe standard betting market dynamics and to predict expected betting volumes conditional on match characteristics. Deviations from these expectations can then be analysed using outlier detection techniques to identify potentially suspicious betting activities. The results demonstrate how statistical modelling can contribute to the early identification of unusual betting patterns, thereby supporting integrity assurance in in-play sports betting markets.

\end{abstract}

\section{Introduction}
\label{sec:introductio}
Match-fixing poses a serious threat to the integrity of sports competitions. It undermines public trust and diminishes fan engagement, thereby threatening the commercial viability of clubs, leagues, and associated industries. It also compromises the role of sport as a vehicle for fair play, community cohesion, and youth development. Match-fixing becomes lucrative and more likely when fixers can profit financially. The proliferation of global sports betting markets is hence accompanied by an increase in opportunities for profitable match-fixing, due to the ease of access and high liquidity, which reduces the chances of detection. If match-fixing were to become a regular and undeniable occurrence, it would further affect public trust and discourage fair athletes \citep{breuer2018palgrave}. This underscores the need for sophisticated, data-driven systems to detect and prevent betting-related corruption, thereby securing the integrity of sports.

Football is the frontrunner in terms of betting turnover worldwide \citep{statista2025} and also the sport with the highest number of suspected match-fixing incidents reported by the industry: In their 2024 annual integrity report, Sportradar flagged 721 suspicious football matches \citep{sportradar2025}. While match-fixing is often associated with individual matches, some scandals even span multiple matches. Notably, the Italian Serie~B has been repeatedly implicated, with the 2011–12 scandal involving players such as Giuseppe Signori and Cristiano Doni, leading to sanctions for clubs and individuals. More recently, the 2015 ``Dirty Soccer'' investigation revealed that Catania Calcio's management orchestrated the fixing of five matches to avoid relegation, resulting in arrests and league penalties \citep{costa2018globalised}. In 2025, a betting scandal in Turkey involving illegal sports betting and referees, club presidents, and footballers was revealed \citep{kocsoy2026determinants}. These incidents highlight the ongoing challenges in safeguarding the integrity of football competitions. 

A major difficulty in assessing fraud detection models lies in the limited availability of verified ground-truth data on match-fixing. However, given the incidents in the Italian Serie~B, this league appears to be a natural platform for testing candidate models for fraud detection and for investigating potentially interesting anomaly patterns in betting behaviour. Although automated data-driven corruption-detection systems can clearly be valuable tools for identifying irregularities and hence protecting the sport, the corresponding literature to date is relatively scarce, given the market size. It involves data analyses and outlier detection methods applied to, e.g.\ basketball \citep{wolfers2006point}, sumo wrestling \citep{duggan2002winning}, and football \citep{reade2013using, otting2018integrating}. However, the existing literature has mostly focused on pre-match markets, despite the fact that the in-play (live) betting market today accounts for about 40\% of the total betting volume and is particularly relevant when examining suspicious matches \citep{annacitvityreport2025}. To comprehensively monitor irregular betting activity, there is thus a pressing need to develop fraud warning systems for the in-play betting market as well. Furthermore, the high resolution and, consequently, the granularity of betting activity surrounding major in-play events may offer additional insights and support fraud detection that would otherwise remain hidden. For example, unusually high betting volumes might be placed on relatively unlikely outcomes shortly before the end of matches -- so as to maximise profits as the odds increase with time running out -- which were manipulated to ensure that outcome.

We extend previous research by analysing in-play betting data to investigate potentially fraudulent betting behaviour in the in-play betting market. Specifically, this paper seeks to identify potentially suspicious deviations from typical patterns in betting market activity during football matches. This requires two steps: First, the dynamics of betting markets under regular conditions need to be understood. This provides the foundation for predicting expected betting volumes. Second, we compare these predictions with the actually observed stakes to detect anomalies. Outlier detection techniques are then applied to flag potentially suspicious matches or situations that may indicate manipulation. To this end, we consider high-resolution betting data from the 2018/19--2020/21 seasons of the Italian Serie~B, provided by a major European bookmaker. The dataset contains second-by-second records of betting volume for potential match outcomes, along with the bookmaker's in-play odds. Following approaches used in general financial market analyses (see, e.g.\ \citealp{barra2017joint}\\ \citep{warne2017marginalized}), we employ a state-space modelling framework to capture the underlying serial correlation in the market activity level. Controlling for key covariates capturing responses to major events \citep{otting2024demand} and in-play dynamics \citep{michels2023bettors} as well as the anticipation of news in betting markets \citep{winkelmann2025betting} as the expected key drivers of betting actions, we isolate unexplained large positive deviations from the model-implied expected betting volumes as the instances of potential interest. Our study can be seen as a proof-of-concept for the practical benefits of advanced statistical modelling as a tool for flagging suspicious betting volumes in in-play betting.

\section{Characteristics of betting markets}
\label{sec:literature}
Sports betting markets are of high economic relevance, given their gross revenue of approximately \euro{}21 billion in Europe \citep{annacitvityreport2025}. From an economic perspective, they share various similarities with financial markets. In particular, placing a bet is conceptually similar to buying a financial asset \citep{sauer1998economics}. A crucial advantage of betting markets is that the outcome of a bet (the asset's uncertain value) becomes observable at a fixed deadline, typically the end of a match \citep{thaler1988anomalies}. According to the efficient market hypothesis, betting odds offered by bookmakers must contain all available information \citep{eugene1970efficient}, thereby eliminating any systematic strategies that could generate returns for bettors. However, bookmakers maximise their profits by ``balancing their books'', i.e.\ they offer odds in response to (expected) bettors' demand, such that they secure profitability. Thus, odds-implied probabilities (corrected for the bookmaker's overround, cf.\ \citealp{winkelmann2024betting}) may deviate from \textit{true} probabilities, challenging the efficient market hypothesis \citep{thaler1988anomalies}.

Potential match-fixing can manifest itself in anomalies, i.e.\ large unexplained deviations from typical patterns in betting markets. Effectively identifying such deviations requires an accurate quantification of the market dynamics. To this end, this section reviews the literature on the structure of both pre-match and in-play betting markets. Furthermore, it provides an overview of recent approaches to detecting betting fraud.

\subsection{Pre-match betting}

Most research on betting markets has focused on pre-match betting (see, e.g.\ \citealp{thaler1988anomalies, dixon2004value, feddersen2017sentiment, deutscher2018betting}), as in-play betting has only recently become more common. However, understanding pre-match market behaviour is also relevant for the in-play market, as in-play betting odds are built on these initial expectations. The majority of the corresponding studies do not focus on fraud detection, but rather on potential market inefficiencies, i.e.\ situations where the odds do not accurately reflect the true probability of an outcome. Nevertheless, these contributions help to understand the dynamics of the betting market and can thus support the development of a fraud detection system.

Market efficiency is often evaluated by testing simple strategies, such as betting on home teams or favourites, with mixed results \citep{hegarty2024comparing}. Some studies find general inefficiencies \citep{constantinou2013profiting, dixon2004value, angelini2019efficiency}, while others highlight specific cases, such as recently promoted teams \citep{deutscher2018betting}, end-of-season matches \citep{borghesi2007late}, or the reduced home advantage during COVID-19 \citep{winkelmann2021bookmakers}. Many of the observed inefficiencies appear to reflect irrational behaviour (biases) rather than fraudulent activity. 

For example, the \textit{favourite-longshot bias} (FLB) refers to some bettors' tendency to overvalue longshots and undervalue favourites \citep{ottaviani2008favorite, snowberg2010explaining}. Several studies confirm the FLB in European football betting markets \citep{cain2000favourite, constantinou2013profiting, buhagiar2018some, angelini2019efficiency}, while others find a reverse FLB, where favourites are overvalued \citep{dixon2004value, gil2007testing, woodland1994market}. \citet{newall2021sports} suggest that the direction of the bias depends on the number of outcomes and the favourite’s implied probability; in addition, the FLB is linked to sentiment and risk preferences \citep{forrest2008sentiment}. Other studies provide evidence that bookmakers undervalue the advantage of playing at home, driven by venue familiarity, reduced travel fatigue, and crowd support \citep{carmichael2005home, carron2005home}, thus offering higher odds for home teams, referred to as \textit{home bias} \citep{forrest2008sentiment, angelini2017parx}. The \textit{sentiment bias} suggests that bettors favour well-known or popular teams, but the evidence in the literature is again mixed \citep{feddersen2017sentiment,forrest2008say}. 

In a comprehensive study, \citet{winkelmann2024betting} analyse common pre-match biases for the top five European football leagues and several seasons. They consider their empirical results in light of a simulation-based analysis, demonstrating that the frequency of inefficiencies is barely higher than what would be expected in a fully efficient market by chance alone. This leads to the conclusion that inefficiencies in pre-match betting markets are, contrary to the impression given by previous literature, neither persistent nor systematic across leagues.

\subsection{In-play betting}

While pre-match odds are relatively stable and mainly respond to injuries or line‑up changes, in-play betting is highly dynamic: beyond pre-match expectations, odds and stakes continuously respond to evolving match dynamics and events, such as goals, red cards, or momentum shifts. While the share of revenue from in-match betting, relative to the total volume of sports betting, fluctuates around 40\% \citep{annacitvityreport2025}, there is still very little publicly available academic research on in-play betting, with most of the existing literature considering the case of betting exchanges. For the latter, \citet{choi2014role} find that bettors underreact to moderately surprising events but overreact to major ones, with the effects vanishing within minutes. Further studies provide mixed evidence regarding market efficiency \citep{croxson2014information,gil2007testing}. In their mispricing detection framework, \citet{angelini2022informational} test efficiency and assess behavioural biases in in-play betting exchange data. Using first goals as events of major news, they identify pre-match and in-play inefficiencies to positively correlate with the level of surprise, driven by biases such as the reverse FLB.

The recent availability of in-match betting data from bookmakers, including both odds offered and stakes placed by bettors, allows for increasingly comprehensive analyses of in-play betting markets. \citet{michels2023bettors} find that betting decisions reflect both pre-match expectations and in-match team strength, with bettors placing more money on matches exhibiting higher outcome uncertainty, particularly in the final 30 minutes of a match \citep{otting2024demand}. Research on goals as a major in-match event with its decisive impact in the low-scoring sport of football suggests that neither bookmakers nor bettors anticipate their occurrence, indicating that unusual odds movements or substantially elevated stakes placed shortly before a team scores may indicate fraudulent behaviour \citep{winkelmann2025betting}. In contrast, market activity does increase after major news, especially after surprising goals \citep{otting2024demand}. This is consistent with the \textit{cognitive bias} in psychology and has been observed across different sports \citep{durand2021behavioral}.

\subsection{Fraud detection in pre-match and in-play betting}

The substantial growth of the online and, in particular, the in-play betting market has fundamentally changed the environment in which match-fixing can occur, offering increasingly many avenues for manipulation and complicating its detection, as betting activity is globally distributed and no unified global monitoring framework exists. Match-fixing occurs more frequently when corrupt behaviour appears justifiable or attractive to protagonists, for example, due to dissatisfaction with low wages or in matches with little at stake, typically towards the end of a season. As a consequence, second- or third-tier leagues, within which players and referees receive comparatively low compensation, while the betting markets are still highly liquid, are especially prone to corruption (for broader discussions of the motives behind match-fixing and documented cases in professional football, see \citealp{forrest2008say,forrest2012threat}). On the referee side, anomalies such as systematically higher betting volumes for matches officiated by certain referees have been documented \citep{deutscher2025match}.

Despite these risk factors outlined above, detecting suspicious activity remains challenging. Unusual betting patterns or poor on-field performance may simply result from players conserving energy in low-stakes matches \citep{forrest2008say}. The detection of fraud in in-play betting is even more complicated due to short time windows within which actions occur, strongly fluctuating odds, and higher betting limits \citep{oddsmatchfix2015}. Recent approaches to fraud detection increasingly rely on data analytics and artificial intelligence. Systems such as Sportradar’s \textit{Universal Fraud Detection System} (UFDS) compare in-play market odds with model-based forecasts \citep{sportradar2024}. In the absence of manipulation, both should closely align as all information is publicly available \citep{reade2013using}. Forecasting methods employed include econometric models \citep{reade2013using}, Bayesian networks \citep{razali2017predicting}, Bayesian inference with rule-based reasoning and time-series inputs \citep{min2008compound}, and forensic statistics for in-play forecasting \citep{forrest2019using}. Other studies rely on machine learning approaches, such as deep neural networks \citep{anfilets2020deep}, long short-term memory networks (LSTMs) models with attention mechanisms \citep{zhang2022sports}, or a combination of different methods \citep{kim2024ai}. Most of the existing literature focuses exclusively on betting odds when detecting market anomalies. However, \citet{otting2018integrating} show that including betting volumes can enhance model accuracy, as high liquidity can stabilise odds and conceal manipulation. The authors combine odds and volumes and apply outlier detection using distributional regression on pre-match Serie B data, successfully identifying suspicious matches in seasons with known fixes. Their findings suggest that continuous monitoring of both odds and volumes can improve match-fixing detection, forming the basis for our study.

\section{Data}
\label{sec:data}
Our dataset is provided by one of Europe's largest bookmakers and contains in-play betting records from three consecutive seasons of the Italian Serie~B (2018/19--2020/21). The sample covers 1,097 matches involving 32 different teams. Four matches were excluded due to missing bookmaker data, and one further match was omitted because it was interrupted by fog and resumed on the following day. A football match consists of two halves of 45 minutes each, typically extended by a few minutes of injury time at the end of each half. For each match, our dataset includes information on match-specific characteristics, bookmaker odds, and betting activity. The original data were recorded at a very high resolution of 1~Hz. To reduce volatility in betting stakes, thereby facilitating the statistical modelling of the betting dynamics, we aggregate observations into one-minute intervals (cf.\ \citealp{michels2023bettors,winkelmann2025betting}). To avoid the need to deal with noise and often extreme covariate values (e.g.\ for the Gini coefficient) during the very last minutes of a match, we exclude observations from the injury time after the second half.

We are specifically interested in bets placed on the home and away teams, as betting on draws is generally unpopular in football -- in fact, in our dataset, more than 80\% of the stakes are placed on one of the two teams -- and, more importantly, draws can be expected to be less prone to fraud, as this would typically require the coordinated cooperation of both teams. Therefore, we mirror the dataset and include each observation twice, once from the home team's perspective and once from the away team's perspective. This results in a total of 235,882 observations. To comply with confidentiality requirements, bettors' stakes are scaled by a fixed constant, ensuring that relative comparisons across observations remain valid. Stakes can only be placed when the betting market is open. We define an open market as the possibility to bet on all three outcomes. Bookmakers close the market right after major events, such as goals and red cards, and towards the end of a match if its outcome is effectively decided. We capture this information by the variable \textit{open} taking the value 1 if the market is open. We observe a closed market ($\textit{open} = 0$) for 9.8\% of our observations. We are mainly interested in the stakes placed on a certain team within a given minute, denoted by the variable \textit{stake\_team}. In our sample, stakes range from 0 to 257.03 and are highly right-skewed (median: 0.37, mean: 1.19; cf.\ Table~\ref{tab:sum_stats}), with no stakes placed on a team despite an open market in 9.4\% of observations. To facilitate the modelling in case of positive stakes, we therefore use the logarithm of these stakes in our models in Section~\ref{sec:stakes} (see the boxplots in Figure~\ref{fig:log_stakes} in Appendix~\ref{sec:app_stakeslog} for a visualisation of stakes).

\begin{figure}[htp!]
    \centering
    \includegraphics[width=0.7\linewidth]{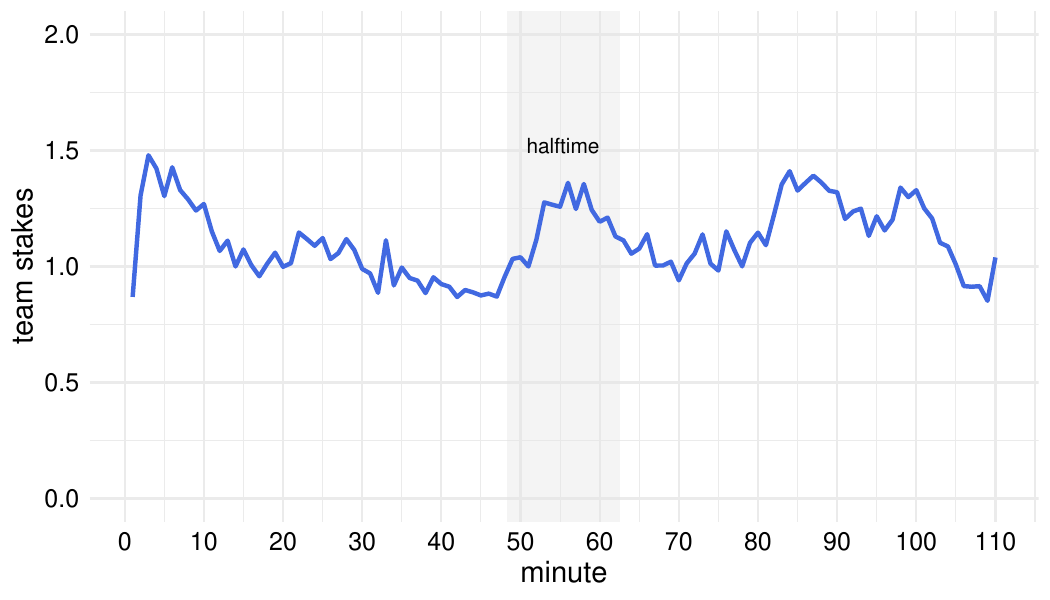}
    \caption{Average stakes placed per minute. The average halftime is highlighted in grey. To avoid misleading impressions induced by a small number of matches with very long injury times, we display average stakes only until 110 minutes after kick-off.}
    \label{fig:stakes_minute}
\end{figure}

The dynamics of betting activity are affected by various drivers, which we quantify using both pre-match and in-match variables. Betting stakes vary strongly with the teams involved in a match.
The average stakes per minute placed on the different teams in our dataset, denoted by \textit{stake\_avg\_team}, are illustrated in Figure~\ref{fig:teamsavgstakes} in Appendix~\ref{sec:app_teamstakes} (for the total stakes placed on each team over the whole observation period, see Figure~\ref{fig:teamstotstakes} in Appendix~\ref{sec:app_totstakes}). The distribution is again right-skewed, with a mean of 1.49 and a median of 1.18. Hellas Verona and US Palermo appear as outliers with exceptionally high average stakes. 
In addition, stakes vary with the \textit{minute} of the match: Figure~\ref{fig:stakes_minute} displays average stakes placed as a function of the time elapsed since kick-off. On average, the highest stakes are placed during the initial 10 minutes of a match. They again increase during halftime and late in the second half.

Generally, pre-match characteristics such as team strength and squad changes are captured by the implied winning probabilities, derived from pre-match odds offered by the bookmaker (cf.\ \citealp{winkelmann2024betting}). We find that the pre-match implied probabilities of teams, denoted by \textit{improbteam\_start}, vary between 7.69\% and 73.85\% (mean 34.15\%). These pre-match probabilities are higher for home teams (42.15\%) than for away teams (28.46\%), reflecting the well-known home advantage. 

In-match implied probabilities, and likely also betting activity, are affected by the score difference, by major events such as goals and red cards, and by the time elapsed since kick-off \citep{winkelmann2025betting}. The number of total goals in our dataset ranges from 0 to 8 (mean: 2.32). Whether a match is effectively decided is mainly determined by the difference in the score between both teams, denoted by \textit{scorediff\_team} from the perspective of the team under consideration. A positive score difference indicates that this team is in the lead. Figure~\ref{fig:volume_scorediff} in Appendix~\ref{sec:app_stakesscorediff} displays the distribution of stakes placed depending on the score difference. The magnitude of in-play shifts in winning probabilities depends on the level of surprise associated with a goal. We follow the definition of \citet{otting2024demand} and denote surprising goals as those scored by teams with a pre-match odds-implied winning probability of at most 0.25. Analogously, slightly surprising (unsurprising) goals are defined as those scored by teams with a pre-match implied winning probability between 0.25 and 0.5 (above 0.5). 

Bettors may react not only to goals themselves but also to major goal-scoring opportunities that potentially shift the dynamics of the match \citep{winkelmann2025betting}. We therefore consider the expected (number of) goals retrieved from \url{https://fbref.com} from the start of the match until a given minute under consideration as an additional covariate. We denote the difference in expected goals between the team under consideration and its opponent as \textit{xg\_diff}, taking a maximum value of 4.78. In addition to goals and goal-scoring opportunities, we consider the variable \textit{redcard\_team}, which denotes the number of red cards issued to the team, whereas the variable \textit{redcard\_opponent} indicates the number of red cards issued to its opponent. Within our dataset, 231 red cards were issued, with at most two within a single match. To measure uncertainty about the match outcome, we calculate the (normalised) Gini coefficient across implied probabilities for the three potential outcomes. The variable \textit{Gini} provides its current value in a given minute. The Gini coefficient ranges from 0 to 1, with lower values indicating greater uncertainty about the match outcome and potentially attracting bettors to place higher stakes.

\begin{table}[htp!]
\centering
\caption{Summary statistics of the key variables considered in our model.}
\label{tab:sum_stats}

\sisetup{
  round-mode          = places,
  round-precision     = 2,
  table-number-alignment = center
}

\begin{tabular}{
  l
  S[table-format=4.2]
  S[table-format=3.2]
  S[table-format=3.2]
  S[table-format=4.2]
  S[table-format=3.2]
}
\toprule
variable & {mean} & {sd} & {min} & {max} & {median} \\
\midrule
stake\_team      & 1.19 & 3.18 & 0  & 257.03 & 0.37 \\
stake\_avg\_team      & 1.42 & 0.9 & 0.51  & 6.6 & 1.23 \\
improbteam\_start     & 0.35 & 0.12 & 0.08  & 0.74 & 0.34 \\
redcard\_team      & 0.03 & 0.17 & 0  & 2 & 0 \\
redcard\_opponent      & 0.03 & 0.17 & 0  & 2 & 0 \\
scorediff\_team   & 0   & 1.04   & -7    & 7       & 0      \\
xg\_diff      & 0 & 0.71 & -4.78  & 4.78 & 0 \\
Gini      & 0.45 & 0.28 & 0  & 0.99 & 0.42 \\
\bottomrule
\end{tabular}
\end{table}

Table~\ref{tab:sum_stats} presents summary statistics for the key variables to be included in our model. Additionally, we consider the situation of no stakes being placed on a team. We observe this situation more frequently when the Gini coefficient is higher (correlation coefficient: 0.41), i.e.\ when the outcome of the match is more certain, and later in the match (correlation coefficient with the minute of the match: 0.23).

\section{Modelling betting market dynamics}
\label{sec:stakes}
The data to be modelled are the sequences of aggregated stakes placed by bettors per minute. These sequences exhibit fairly strong serial correlation, induced by the underlying market dynamics, with matches generally going through varying levels of excitement -- and thus betting activity. To capture this dependence of the stakes on the not directly observable market activity level, we employ state-space models (SSMs) to model stakes placed on a team at a given point in a match, conditional on the underlying market activity level. This approach is analogous to stochastic volatility modelling in financial markets, where the share returns are assumed to be driven by the underlying latent nervousness of the market (see, e.g.\ \citealp{taylor1994modeling,abanto2017maximum}), and has already successfully been transferred to the context of betting markets \citep{michels2023bettors,otting2024demand}. In the following, we outline the modelling framework and describe how to conduct statistical inference for this model. We then report results for models of increasing complexity and proceed to conduct outlier detection based on the final model.

\subsection{Modelling framework and parameter estimation}
\label{sec:framework}

An SSM comprises an observed state-dependent process $\{y_t\}$ -- these will be the sequences of stakes placed -- and an unobserved state process $\{s_t\}$, a proxy for the overall market activity level. We are interested in the absolute stakes $y_{ijt}$ placed on team $i$ in match $j$ and minute $t$. For ease of notation, we will omit the subscripts $i$ and $j$ in the following. Given the strong skewness of the stakes, we consider the logarithm of these stakes to facilitate the modelling. We then use a hurdle model, with a point mass for the case where no stakes are placed in a given minute, i.e.\
$$
\Pr (y_t = 0) = \pi_t, \quad \text{hence} \quad \Pr (y_t > 0) = 1-\pi_t,
$$
fixing $\pi_t = 1$ for all minutes where the market is closed.
For the log-transformed positive stakes, we assume
$$\log (y_t) \mid y_t > 0 \;=\; \mu_t + \sigma \sqrt{\tfrac{\nu-2}{\nu}}\, \xi_t, \qquad \xi_t \sim t_\nu,$$
i.e.\ a location-scale Student's t-distribution with $\nu > 2$ degrees of freedom, parameterised such that $\text{Var}\bigl(\log(y_t) \mid y_t > 0\bigr) = \sigma^2$. 
In other words, the distribution assumed for $y_t$ is a two-component mixture of a point mass on zero with a location-scale t-distribution, the latter on the log scale. Compared to the normal distribution, the t-distribution places more probability mass both close to its centre and in the tails, with the normal distribution as the limiting case for $\nu \to \infty$. An analysis of (pseudo-)residuals of an initial model specification with normally distributed errors indicated the necessity to account for such heavy tails (for details, see Section~\ref{sec:checking}). The standardisation ensures that $\sigma$ retains its interpretation as the standard deviation of the log-transformed stakes. The mean of the log-transformed stakes is modelled as $\mu_t = \eta_t + s_t$, with a linear predictor $\eta_t = \mathbf{x'}_t^{(1)}\boldsymbol{\beta}$ comprising the relevant covariates and $s_t$ the latent market activity level. To capture the time-varying levels of excitement during a match, the latter is assumed to be a covariate-driven AR(1) process,
$$
s_t = \phi s_{t-1} + \mathbf{x'}_t^{(2)}\boldsymbol{\omega}+ \sigma_s \epsilon_t,
$$
where $\lvert \phi \rvert < 1$ is the persistence parameter, $\epsilon_t \sim \mathcal{N}(0,1)$, $\sigma_s > 0$, and covariates affecting the state process $\mathbf{x'}_t^{(2)}$. The dependence structure of the SSM is displayed in Figure~\ref{fig:ssm}.

\begin{figure}[!ht]
    \centering
    \scalebox{0.65}{
    \begin{tikzpicture}
    \node[circle,draw=black, fill=gray!5, inner sep=0pt, minimum size=50pt, font=\LARGE] (A) at (-2, -2.8) {...};
    \node[circle,draw=black, fill=gray!5, inner sep=0pt, minimum size=50pt, font=\LARGE] (B) at (1, -2.8) {$s_{t-1}$};
    \node[circle,draw=black, fill=gray!5, inner sep=0pt, minimum size=50pt, font=\LARGE] (C) at (4, -2.8) {$s_{t}$};
    \node[circle,draw=black, fill=gray!5, inner sep=0pt, minimum size=50pt, font=\LARGE] (D) at (7, -2.8) {$s_{t+1}$};
    \node[circle,draw=black, fill=gray!5, inner sep=0pt, minimum size=50pt, font=\LARGE] (E) at (10, -2.8) {...};
    
    \node[circle,draw=black, fill=gray!5, inner sep=0pt, minimum size=50pt, font=\LARGE] (F) at (1, 0) {$y_{t-1}$};
    \node[circle,draw=black, fill=gray!5, inner sep=0pt, minimum size=50pt, font=\LARGE] (G) at (4, 0) {$y_{t}$};
    \node[circle,draw=black, fill=gray!5, inner sep=0pt, minimum size=50pt, font=\LARGE] (H) at (7, 0) {$y_{t+1}$};

    \node[circle,draw=black, fill=gray!5, inner sep=0pt, minimum size=50pt, font=\LARGE] (I) at (1, 2.8) {$\mathbf{x}^{(1)}_{t-1}$};
    \node[circle,draw=black, fill=gray!5, inner sep=0pt, minimum size=50pt, font=\LARGE] (J) at (4, 2.8) {$\mathbf{x}^{(1)}_{t}$};
    \node[circle,draw=black, fill=gray!5, inner sep=0pt, minimum size=50pt, font=\LARGE] (K) at (7, 2.8) {$\mathbf{x}^{(1)}_{t+1}$};

    \node[circle,draw=black, fill=gray!5, inner sep=0pt, minimum size=50pt, font=\LARGE] (L) at (1, -5.4) {$\mathbf{x}^{(2)}_{t-1}$};
    \node[circle,draw=black, fill=gray!5, inner sep=0pt, minimum size=50pt, font=\LARGE] (M) at (4, -5.4) {$\mathbf{x}^{(2)}_{t}$};
    \node[circle,draw=black, fill=gray!5, inner sep=0pt, minimum size=50pt, font=\LARGE] (N) at (7, -5.4) {$\mathbf{x}^{(2)}_{t+1}$};

    \draw[-{Latex[scale=2]}] (A)--(B);
    \draw[-{Latex[scale=2]}] (B)--(C);
    \draw[-{Latex[scale=2]}] (C)--(D);
    \draw[-{Latex[scale=2]}] (D)--(E);
    \draw[-{Latex[scale=2]}] (B)--(F);
    \draw[-{Latex[scale=2]}] (C)--(G);
    \draw[-{Latex[scale=2]}] (D)--(H);
    \draw[-{Latex[scale=2]}] (I)--(F);
    \draw[-{Latex[scale=2]}] (J)--(G);
    \draw[-{Latex[scale=2]}] (K)--(H);
    \draw[-{Latex[scale=2]}] (L)--(B);
    \draw[-{Latex[scale=2]}] (M)--(C);
    \draw[-{Latex[scale=2]}] (N)--(D);

    \draw [decorate, decoration={brace, amplitude=8pt}]
      ($(F.east)+(10,0.8)$) -- ($(F.east)+(10,-0.8)$)
      node [midway, right=8pt] {\LARGE \quad observed (state-dependent process)};

    \draw [decorate, decoration={brace, amplitude=8pt}]
      ($(B.east)+(10,0.8)$) -- ($(B.east)+(10,-0.8)$)
      node [midway, right=8pt] {\LARGE \quad latent (state process)};

    \draw [decorate, decoration={brace, amplitude=8pt}]
      ($(I.east)+(10,0.8)$) -- ($(I.east)+(10,-0.8)$)
      node [midway, right=8pt] {\LARGE\quad\shortstack[l]{covariates for the\\state-dependent process}};

    \draw [decorate, decoration={brace, amplitude=8pt}]
      ($(L.east)+(10,0.8)$) -- ($(L.east)+(10,-0.8)$)
      node [midway, right=8pt] {\LARGE \quad  covariates for the state process};

    \end{tikzpicture}}
\caption{Dependence structure of the SSM with latent state $s_t$ satisfying the Markov property, covariates $\mathbf{x}^{(1)}_t$ affecting the state-dependent process, covariates $\mathbf{x}^{(2)}_t$ affecting the state process, and response variable $y_t$ giving the log-transformed stakes.}
\label{fig:ssm}
\end{figure}

The model is estimated via maximum likelihood using the techniques first introduced by \citet{kitagawa1987non}, where the state space is very finely discretised to address the otherwise analytically intractable likelihood. Integrating over all possible state sequences and exploiting the dependence structure shown in Figure \ref{fig:ssm} (cf.\ \citealp{ZML}), the exact likelihood is obtained as follows:
\begin{equation}
\mathcal{L}_T (\bm{\Theta}) = \int\cdots\int f(s_1)f(y_1|s_1,\mathbf{x}_1^{(1)})\prod\limits_{t=2}^{T} f(s_t|s_{t-1},\mathbf{x}_t^{(2)})f(y_t|s_t,\mathbf{x}_t^{(1)})ds_T\ldots ds_1,\\
\label{eq:SSMllk}
\end{equation}
with the parameter vector $\Theta$ comprising all regression coefficients as well as the error variances, the persistence parameter of the AR(1) process, and the degrees of freedom $\nu$. To very closely approximate this likelihood, we then apply numerical integration via a simple midpoint rule, finely discretising the state space into $m$ intervals $B_i=(b_{i-1},b_i)$ with length $h=(b_m-b_0)/m$ and midpoints $b_i^*=0.5(b_i+b_{i-1})$, with a large $m$ (typically around 100; cf.\ \citealp{langrock2012some,mews2024maximum,winkelmann2025momentum}). This yields the approximation
\begin{equation}
{\mathcal{L}}_T (\bm{\Theta}) \approx h^T\sum\limits_{i_1=1}^m\cdots\sum\limits_{i_{T}=1}^m f(b_{i_1}^*)f(y_1|b_{i_1}^*,\mathbf{x}_1^{(1)})\prod\limits_{t=2}^{T}f(b_{i_t}^*|b_{i_{t-1}}^*,\mathbf{x}_t^{(2)})f(y_t|b_{i_t}^*,\mathbf{x}_t^{(1)}).
\label{eq:llk}
\end{equation}
The numerical integration corresponds to a close approximation of the SSM by a hidden Markov model (HMM) with a large number of states $m$ and highly structured state transition dynamics. The computational cost of evaluating the likelihood of the approximating HMM can therefore be dramatically reduced, from the order $\mathcal{O}(Tm^T)$ to the order $\mathcal{O}(Tm^2)$, by making use of the recursive forward algorithm, a standard HMM tool (\citealp{ZML}, Chapter~11). For this, we summarise the approximate probabilities of switching from interval $i$ to interval $j$, $\gamma_{ij} = h f(s_t = b^*_j | s_{t-1} = b_i^*,\mathbf{x}_t^{(2)})$, $i,j=1,\ldots, m$, in the transition matrix $\Gamma$. This matrix is fully determined by the parameters $\phi$ and $\sigma_s$ as well as the regression coefficients $\boldsymbol{\omega}$. We further define the $m \times m$ diagonal matrix $\mathbf{P}(y_t)$ with the state-dependent densities $f(y_t | s_t = b_i,\mathbf{x}_t^{(1)})$, $i=1,\ldots,m$, from the above hurdle model on the diagonal. Application of the forward algorithm then allows us to rewrite the (approximate) likelihood (\ref{eq:llk}) as the matrix product
\begin{equation*}
{\mathcal{L}}_T (\bm{\Theta}) \approx \bm{\delta} \mathbf{P}(y_1) \bm{\Gamma} \mathbf{P}(y_2) \ldots \bm{\Gamma} \mathbf{P}(y_{T}) \bm{1},
\end{equation*}
\noindent
with $\bm{1} = (1,\ldots,1)' \in \mathbb{R}^m$ and $\bm{\delta}$ the initial distribution of the AR(1) process.

For the given longitudinal dataset, we assume independence across matches and teams, considering each match from the perspective of each team as a single time series (match-team-series). Consequently, the total likelihood is calculated as the product of individual likelihoods of the above form. To obtain parameter estimates, we numerically maximise the likelihood in Python via the Broyden–Fletcher–Goldfarb–Shanno (BFGS) algorithm, a quasi-Newton method, subject to the usual technical details arising in effectively every numerical optimisation \citep{ZML}.

\subsection{A baseline model}

As a benchmark for the more complex models below, we initially consider a model without any of the covariates described in Section~\ref{sec:data}. The parameter estimates with their 95\%~confidence intervals for this baseline model are given in Table~\ref{tab:results_baseline}. The estimated persistence of $\hat{\phi} = 0.992$ in the latent AR(1) process indicates strong serial correlation in the market activity level. Further justification for the use of the SSM framework is provided by the Akaike Information Criterion (AIC), which shows a dramatic improvement of the fit compared to a simple hurdle model without an underlying state process ($\Delta \text{AIC} = 173,302$). Furthermore, the usage of the Student's t-distribution is verified by $\hat{\nu} = 4.123$, indicating substantially heavier tails than under normality, and an improved AIC ($\Delta \text{AIC} = 12,549$ compared with a normal distribution). The probability of no stakes being placed in a given minute with an open market is $\hat{\pi} = 0.094$ and thus identical to the empirical frequency of 0.094. 

\begin{table}[htp!]
    \centering
    \begin{tabular}{r|cl}
        parameter & estimate & 95\%~confidence interval \\
        \hline
        $\phi$\, & \,\,0.992 & [\,\,0.992;\,\,\,\,0.992] \\
        $\sigma_s$\, & \,\,0.216 & [\,\,0.215;\,\,\,\,0.217] \\
        \hdashline
        $\beta_0$\, & -0.740 & [-0.747;\,\,-0.734] \\
        $\sigma$\, & \,\,0.953 & [\,\,0.950;\,\,\,\,0.955] \\
        $\nu$\, & \,\,4.123 & [\,\,4.109;\,\,\,\,4.136] \\        
        \hdashline
        $\pi$\, & \,\,0.094 & [\,\,0.093;\,\,\,\,0.094] \\
    \end{tabular}
    \caption{Parameter estimates and 95\%~confidence intervals for the baseline model, where $\mu_t= \beta_0 + s_t$ and $s_t$ is a simple AR(1) process.}
    \label{tab:results_baseline}
\end{table}

\subsection{Including covariates in the state-dependent process}

Accurately identifying outliers for fraud detection requires the comparison of the actual stakes placed to their probability distribution given the current market situation. The latter needs to take into account not only the general dynamics of the in-play betting market as in the baseline model above, but also team- and match-specific variables and events that may elevate or reduce expected betting activity. We thus extend our model based on suggestions from the literature and insights from our descriptive analysis. \citet{otting2024demand} show that team strength affects bettors' tendency to place stakes. In our data, the implied winning probability correlates positively with both the expected goal difference (0.49) and the average stakes placed on the team (0.09). Consistent with this pattern, favourites attract larger stakes, with an average stake of 1.98 compared with 1.34 for underdogs. Since simple measures of team strength such as Elo ratings may not fully capture the overall sentiment towards a team -- which however has been shown to have an effect on betting behaviour \citep{forrest2008sentiment} -- we consider the average stakes per minute placed on the team across the whole observation period, denoted as \textit{stake\_avg\_team}, as a measure capturing information on both team strength and sentiment. The inclusion of this covariate ensures that under our model, the actual stakes placed are compared to their overall level (and distribution) for the specific team under consideration. This is crucial, given the strong variation in the stakes placed across teams as shown in Figure~\ref{fig:teamsavgstakes} in Appendix~\ref{sec:app_teamstakes}. 

To further account for bettors' tendency to place higher stakes on teams designated as favourites, we include the pre-match implied probability for the team considered \textit{improbteam\_start}. In contrast to the variable \textit{stake\_avg\_team}, this measure captures the team's relative strength by explicitly accounting for the strength of the opponent. According to \citet{winkelmann2025betting}, a red card issued to the opponent (team) increases (decreases) average stakes placed on the team, leading us to include the variables \textit{redcard\_team} and \textit{redcard\_opponent}, denoting the number of red cards issued to the team and opponent, respectively. This is confirmed in our data by a positive correlation between red cards issued to the away team and stakes placed on the home team (0.17), and vice versa (0.15).

To also account for in-play dynamics, we further include the variable \textit{scorediff\_team} denoting the current score difference from the perspective of the team under consideration. Finally, the descriptive analysis shows that stakes vary with the minute. Specifically, Figure~\ref{fig:stakes_minute} in Section~\ref{sec:data} shows a $\cup$-shaped relation between the minute and the stakes placed outside halftime (where betting activity is also elevated). We thus include a quadratic effect of the \textit{minute} (in line with \citealp{winkelmann2025betting}) as well as a binary variable \textit{halftime}, taking the value 1 if the match is in its halftime. We standardise the variable \textit{minute} by subtracting its mean and dividing by its standard deviation to ensure numerical stability. In preliminary checks, we also tested whether stakes on home teams are higher than those on away teams, beyond what is already captured in pre-match implied probabilities, and whether the average stake placed on the opponent affects stakes on the team under consideration. However, both possible effects were deemed to be insignificant. Overall, we arrive at the following time-varying linear predictor for the mean stakes, on the log-scale:
\vspace{-0.2cm}
\begin{equation*}
    \begin{split}
    \mu_t = \beta_0\; & + \beta_1 \text{stake\_avg\_team}_t + \beta_2 \text{ improbteam\_start}_t + \beta_3 \text{ redcard\_team}_t + \beta_4 \text{ redcard} \\
    & \text{\_opponent}_t +\beta_5 \text{ scorediff\_team}_t + \beta_6 \text{ minute}_t + \beta_7 \text{ minute}_t^2 + \beta_8 \text{ halftime}_t + s_t
    \end{split}
    \label{eq:covariates_mean}
\end{equation*}
Additionally, our descriptive statistics show that the case of no stakes being placed despite an open market arises more frequently when the match is already decided -- as indicated by a high value of the Gini coefficient -- in particular as the match progresses (cf.\ \citealp{otting2024demand}). We therefore model the probability that zero stakes are placed as
\vspace{-0.2cm}
\begin{equation*}
    \pi_t = \text{logit}^{-1}\left( \alpha_0 + \alpha_1 \text{ Gini}_t + \alpha_2 \text{ Gini}_t^2 + \alpha_3 \text{ minute}_t + \alpha_4 \text{ minute}_t \cdot \text{ Gini}_t + \alpha_5 \text{ minute}_t \cdot \text{ Gini}_t^2 \right)
\end{equation*}
\vspace{-1cm}
\begin{table}[htp!]
    \centering
    \scalebox{0.8}{
    \begin{tabular}{rc|cl}
        variable & parameter & estimate & 95\%~confidence interval \\
        \hline
        & $\phi$ & \,\,0.985 & [\,\,0.985;\,\,\,\,0.985] \\
        & $\sigma_s$ & \,\,0.206 & [\,\,0.203;\,\,\,\,0.208] \\
        \hdashline
        \text{intercept} & $\beta_0$ & -2.310 & [-2.377;\,\,-2.242] \\
        \text{stake\_avg\_team} & $\beta_1$ & \,\,0.249 & [\,\,0.223;\,\,\,\,0.276] \\
        \text{improbteam\_start} & $\beta_2$ & \,\,2.582 & [\,\,2.525;\,\,\,\,2.638] \\
        \text{redcard\_team} & $\beta_3$ & -0.369 & [-0.454;\,\,-0.285] \\
        \text{redcard\_opponent} & $\beta_4$ & \,\,1.208 & [\,\,1.139;\,\,\,\,1.277] \\
        \text{scorediff\_team} & $\beta_5$ & \,\,0.382 & [\,\,0.359;\,\,\,\,0.406] \\
        \text{minute} & $\beta_6$ & \,\,0.043 & [\,\,0.025;\,\,\,\,0.061] \\
        $\text{minute}^2$ & $\beta_7$ & \,\,0.028 & [\,\,0.007;\,\,\,\,0.048] \\
        \text{halftime} & $\beta_8$ & \,\,0.197 & [\,\,0.178;\,\,\,\,0.216] \\
        \hdashline
        & $\sigma$ & \,\,0.953 & [\,\,0.948;\,\,\,\,0.958] \\
        & $\nu$ & \,\,4.087 & [\,\,4.049;\,\,\,\,4.125] \\
        \hdashline
        intercept & $\alpha_0$ & -4.818 & [-4.873;\,\,-4.763] \\
        \text{Gini} & $\alpha_1$ & \,\,3.760 & [\,\,3.674;\,\,\,\,3.846] \\
        $\text{Gini}^2$ & $\alpha_2$ & \,\,0.965 & [\,\,0.932;\,\,\,\,0.998] \\
        \text{minute} & $\alpha_3$ & -0.980 & [-1.020;\,\,-0.940] \\
        \text{minute} $\cdot$ $\text{Gini}$ & $\alpha_4$ & \,\,1.764 & [\,\,1.704;\,\,\,\,1.825] \\
        \text{minute} $\cdot$ $\text{Gini}^2$ & $\alpha_5$ & -0.501 & [-0.534;\,\,-0.467] \\
    \end{tabular}}
    \caption{Parameter estimates and 95\%~confidence intervals for the model including covariates in the state-dependent process.}
    \label{tab:results_statedependent}
\end{table}

Table~\ref{tab:results_statedependent} reports the estimated parameter values together with their 95\%~confidence intervals. The fitted state process confirms strong serial correlation in the placement of betting stakes. We further see confirmation of the bettors' tendency to place higher stakes on the pre-match favourite, as indicated by the positive value of $\hat{\beta}_2$. Consistent with the findings of \citet{winkelmann2025betting} for the German Bundesliga, bettors tend to place higher (lower) stakes when the opponent (team) has received a red card, with the effect of a red card issued to the opponent being stronger. In addition, it is confirmed that bettors place higher stakes when the team's in-match winning chances increase (cf.\ $\hat{\beta}_5$). Finally, with respect to the minute of the match considered, we observe the highest stakes at the beginning and the end of a match, with average stakes additionally elevated during the halftime break by approximately 21.8\%.

\subsection{Including covariates in the state process}

Not only the observed betting stakes, but also their underlying market activity level may respond to covariates. \citet{michels2023bettors} already showed that betting activity is strongly affected by in-match dynamics. We thus extend the AR(1) process used for the market activity level to an ARX(1) process, such that the state process of the corresponding SSM is affected by covariates. As scoring opportunities may affect the underlying betting dynamics, we include the difference in expected goals \textit{xg\_diff} as a covariate in the state process. Furthermore, the persistence in betting activity may differ during halftime (see Figure~\ref{fig:stakes_minute} in Section~\ref{sec:data}), such that we include the variable \textit{halftime} also in the state process. Additionally, \citet{otting2024demand} suggest that bettors increase their activity after a goal has been scored, with stronger effects following surprising goals. This effect is expected to fade within a few minutes after the goal. To capture this effect, we use indicator variables $I_w \in \{0,1\}$, taking the value 1 if the most recent goal was surprising ($w = 1$), slightly surprising ($w = 2$), or unsurprising ($w = 3$) -- and all $I_w=0$ if no goal has been scored yet. The variable $\text{goal\_team}_t^{(w)}$ measures the time in minutes since that last goal (set to the value 1 if no such goal has occurred yet). This yields the following specification of the state process:
$$
s_t = \phi s_{t-1} + \sum\limits_{w=1}^3 \omega_w  I_w/\text{goal\_team}_t^{(w)} + \omega_4 \text{ xg\_diff}_t + \omega_5 \text{ halftime}_t + \sigma_s \epsilon_t.
$$

Table~\ref{tab:results_state} provides the results for this full model. Most estimated coefficients associated with the covariates affecting the state-dependent stakes differ only slightly from those presented in Table~\ref{tab:results_statedependent}. In contrast to the German Bundesliga, where \textit{Valuing Actions by Estimating Probabilities} was found to affect the underlying (latent) betting activity \citep{michels2023bettors}, in the Serie B we do not find the match dynamics, measured by the difference in expected goals, to significantly affect the state of the market (cf.\ $\hat{\omega}_4$). Betting activity during halftime is slightly elevated. The results on the reactions to goals are similar to those reported for the German Bundesliga (cf.\ \citealp{otting2024demand}): While surprising goals, i.e.\ those scored by pre-match underdogs, significantly increase the state and hence the market activity level, unsurprising goals, i.e.\ those scored by pre-match favourites, significantly decrease the market activity.

\begin{table}[htp!]
    \centering
    \scalebox{0.8}{
    \begin{tabular}{rc|cl}
        variable & parameter & estimate & 95\%~confidence interval \\
        \hline
        & $\phi$ & \,\,0.985 & [\,\,0.984;\,\,\,\,0.986] \\
        & $\sigma_s$ & \,\,0.197 & [\,\,0.193;\,\,\,\,0.201] \\
        \hdashline
        \text{intercept} & $\beta_0$ & -2.572 & [-2.636;\,\,-2.508] \\
        \text{stake\_avg\_team} & $\beta_1$ & \,\,0.226 & [\,\,0.206;\,\,\,\,0.245] \\
        \text{improbteam\_start} & $\beta_2$ & \,\,3.420 & [\,\,3.297;\,\,\,\,3.542] \\
        \text{redcard\_team} & $\beta_3$ & -0.363 & [-0.451;\,\,-0.275] \\
        \text{redcard\_opponent} & $\beta_4$ & \,\,1.178 & [\,\,1.107;\,\,\,\,1.248] \\
        \text{scorediff\_team} & $\beta_5$ & \,\,0.420 & [\,\,0.397;\,\,\,\,0.443] \\
        \text{minute} & $\beta_6$ & \,\,0.052 & [\,\,0.031;\,\,\,\,0.073] \\
        $\text{minute}^2$ & $\beta_7$ & \,\,0.030 & [\,\,0.017;\,\,\,\,0.043] \\
        \text{halftime} & $\beta_8$ & \,\,0.197 & [\,\,0.179;\,\,\,\,0.215] \\
        \hdashline
        \text{surprising} & $\omega_1$ & \,\,0.274 & [\,\,0.245;\,\,\,\,0.303] \\
        \text{slightly\_surprising} & $\omega_2$ & -0.037 & [-0.053;\,\,-0.021] \\
        \text{unsurprising} & $\omega_3$ & -0.394 & [-0.422;\,\,-0.366] \\
        \text{xg\_diff} & $\omega_4$ & \,\,0.001 & [-0.000;\,\,\,\,0.003] \\
        $\text{halftime}$ & $\omega_5$ & \,\,0.004 & [\,\,0.001;\,\,\,\,0.007] \\

        \hdashline
        & $\sigma$ & \,\,0.955 & [\,\,0.948;\,\,\,\,0.962] \\
        & $\nu$ & \,\,4.103 & [\,\,4.007;\,\,\,\,4.203] \\
        \hdashline
        intercept & $\alpha_0$ & -4.819 & [-4.864;\,\,-4.773] \\
        \text{Gini} & $\alpha_1$ & \,\,3.761 & [\,\,3.559;\,\,\,\,3.963] \\
        $\text{Gini}^2$ & $\alpha_2$ & \,\,0.964 & [\,\,0.745;\,\,\,\,1.184] \\
        \text{minute} & $\alpha_3$ & -0.980 & [-1.030;\,\,-0.930] \\
        \text{minute} $\cdot$ $\text{Gini}$ & $\alpha_4$ & \,\,1.765 & [\,\,1.492;\,\,\,\,2.039] \\
        \text{minute} $\cdot$ $\text{Gini}^2$ & $\alpha_5$ & -0.502 & [-0.776;\,\,-0.227] \\
    \end{tabular}}
    \caption{Parameter estimates and 95\%~confidence intervals for the model including covariates in the state process and in the state-dependent process.}
    \label{tab:results_state}
\end{table}

\subsection{Model diagnostics}
\label{sec:checking}

Performing outlier detection to reliably flag suspicious observations requires an accurate predictive model of betting stakes. We therefore conduct several diagnostic checks to assess the model’s ability to comprehensively capture the market's dynamics. We restrict the model diagnostics to observations for which the market was open. Our hurdle model consists of two components. First, it models the probability of observing a zero stake, $\pi$, as a function of the match minute and the Gini coefficient. Second, conditional on a positive stake, the model uses the SSM described above to explain log-transformed stake sizes. We begin by analysing the latter, since the case of positive stakes is more relevant to our study's aim.

To evaluate the SSM, we recursively compute the model- and data-implied one-step-ahead state probabilities, $p_{t\mid t-1}(i)=\Pr (s_t=i\mid y_{1:t-1},x_{1:t})$, with the states $i$ corresponding to the intervals $B_i$ from the discretisation of the market activity level described in Section \ref{sec:framework}. From this, we obtain the one-step-ahead predictive distribution as
$$
f\bigl( \log (y_t) \mid y_{1:t-1}, x_{1:t}\bigr)
= \sum_{i=1}^m p_{t \mid t-1}(i)\, f \bigl( \log (y_t) \mid s_t=i, x_t \bigr).
$$
Let $F_t \bigl(( \log (y) \bigr)=\Pr \bigl( \log (Y_t) \le y\mid y_{1:t-1},x_{1:t}\bigr)$ denote the corresponding one-step-ahead predictive CDF (a mixture of Student-$t$ distributions). We compute probability integral transform (PIT) values $u_t=F_t(\log (y_t))$ and form quantile pseudo-residuals $r_t=\Phi^{-1}(u_t)$. Under a correct specification -- i.e.\ in case the statistical model closely mimics the real data-generating process -- the pseudo-residuals should be approximately i.i.d.\ standard normally distributed. The most extreme values of $r_t$ are the ones which are the least likely to have occurred under the given model. For the subsequent outlier detection (see Section~\ref{sec:outlier}), we are mainly interested in particularly large positive values, as these may signal suspiciously high betting activity.

\begin{figure}[htp!]
    \centering
    \begin{subfigure}[t]{0.48\linewidth}
        \centering
        \includegraphics[width=\linewidth]{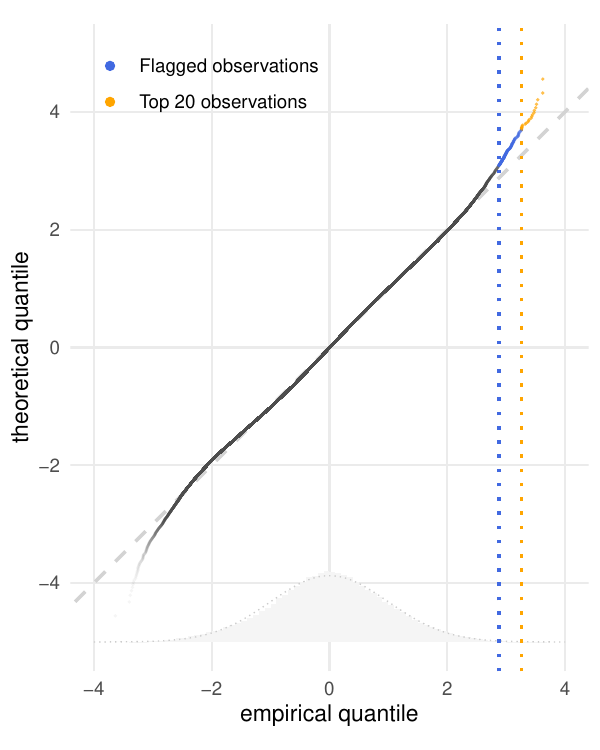}
        \caption{Q--Q plot of pseudo-residuals associated with positive betting stakes. The dashed grey 45-degree line indicates the expected alignment between empirical and theoretical quantiles if the pseudo-residuals were standard-normally distributed. The empirical distribution of the pseudo-residuals is shown along the horizontal axis, with the N(0,1) density overlaid as a dotted line. Points corresponding to stakes exceeding the 99.8\% quantile (flagged observations) and the 99.94\% quantile (top 20 observations) of their respective one-step-ahead predictive distributions are coloured.}
        \label{fig:residuals_qq}
    \end{subfigure}\hfill
    \begin{subfigure}[t]{0.48\linewidth}
        \centering
        \includegraphics[width=\linewidth]{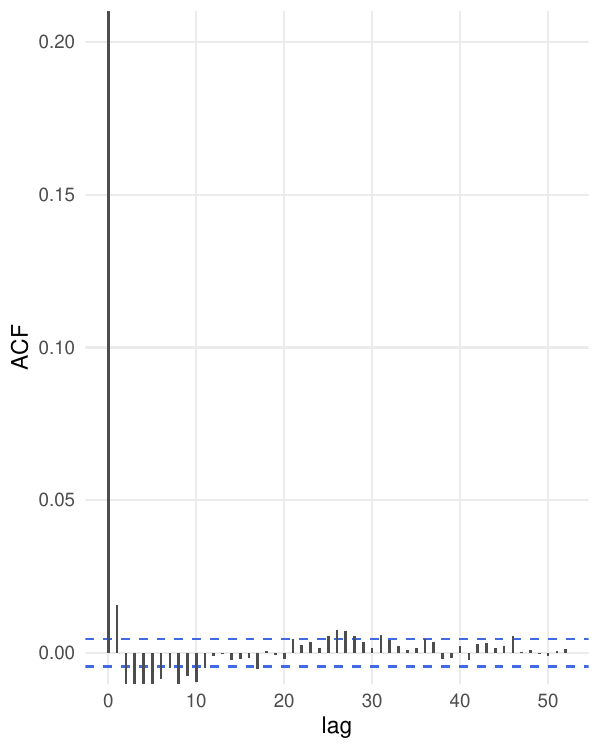}
        \caption{
        Autocorrelation function (ACF) of the pseudo-residuals for positive betting stakes in our dataset. Dashed lines indicate the 95\%~confidence interval under independence over time. The vertical axis is truncated at 0.2. The value at lag 0 is 1 by construction.}
        \label{fig:residuals_acf}
    \end{subfigure}
    \caption{Model diagnostics for the state-space model (SSM) based on pseudo-residuals.}
    \label{fig:ssm_diagnostics}
\end{figure}

Figure~\ref{fig:residuals_qq} shows a Q--Q plot comparing theoretical and empirical quantiles (pseudo-residuals), which indicates an overall reasonable model fit. Notable deviations from the standard normal distribution -- which would be expected under a perfect model specification -- arise only in the extreme tails, specifically concerning the smallest as well as the largest 0.5\% of all observations. For our purposes, these deviations are relevant only in the upper tail, where observed betting activity exceeds what the model predicts. We therefore identify the 195 largest observations as those exceeding the 99.8\%~quantile of the predictive distribution, with particular focus on the top 20. We will examine these outliers more closely in the outlier analysis (see Section~\ref{sec:outlier}). Figure~\ref{fig:residuals_acf} indicates very weak serial dependence in the pseudo-residuals, suggesting that the SSM does a reasonable job at capturing the time dependence between consecutive stakes.

As a check of our model's predictive performance, we further simulate the one-step-ahead predictive distribution of (log-)stakes and compare it with the empirical series. For each time point $t$, we consider the one-step-ahead state probabilities $p_{t|t-1}(i)$ 
and sample 10,000 interval midpoints from this state distribution, conditional on which we compute the predicted time-varying mean $\hat{\mu}$ using the estimated model parameters (see Table~\ref{tab:results_state}).
We then sample innovations  $t_{\hat{\nu}}$ from the fitted Student-$t$ distribution and obtain draws for predicted log-stakes as 
$$
\log (y_{t}^{\text{pred}})
= \hat{\mu}_{t} + \hat{\sigma}\sqrt{\frac{\hat{\nu}-2}{\hat{\nu}}}\,\xi_{t}.
$$

\begin{figure}[htp!]
    \centering
    \includegraphics[width=0.8\linewidth]{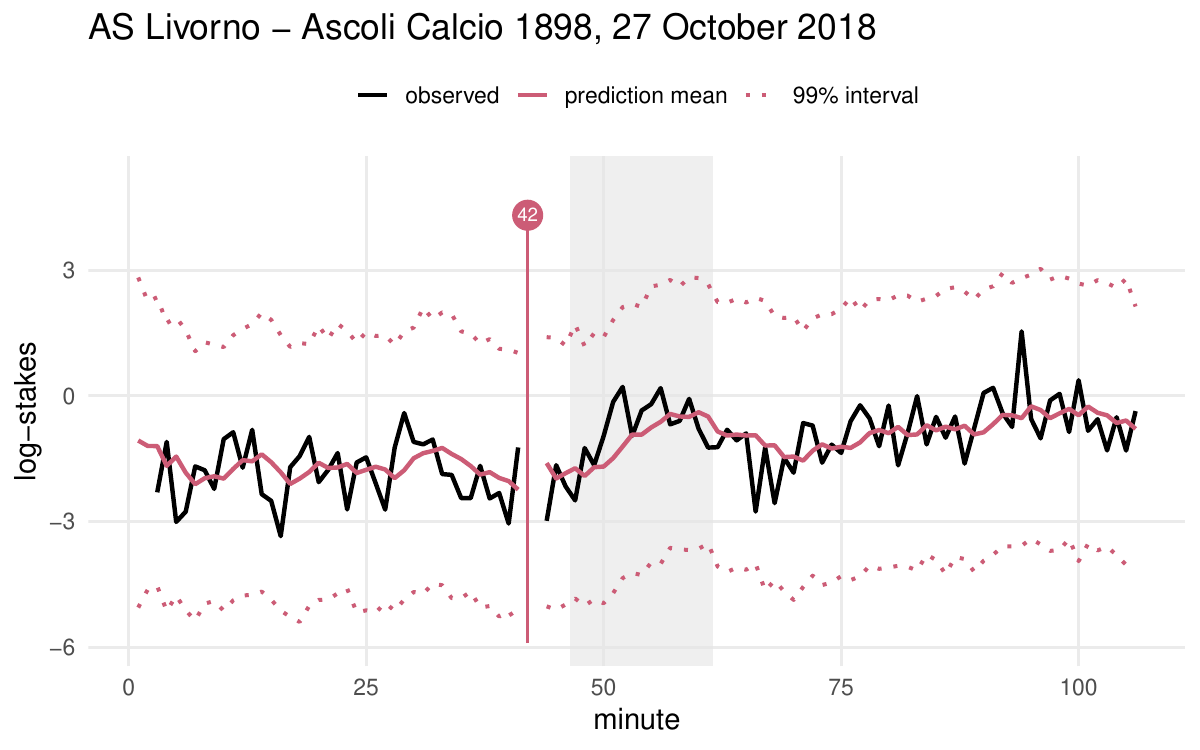}
    \caption{One-step-ahead predictions together with corresponding intervals covering the central 99\% of all predictions for log-stakes placed on AS Livorno in their home match against Ascoli Calcio on 27 October 2018. The vertical line indicates the 1-0 goal scored by AS Livorno in minute 42. The grey-shaded area corresponds to halftime.}
    \label{fig:match_pred}
\end{figure}

Figure~\ref{fig:match_pred} shows the predicted mean and the 99\% predictive interval for the log-stakes placed on AS Livorno in their home match against Ascoli Calcio on 27 October 2018. During the first 40 minutes, average stakes were slightly below the predicted mean. After the opening goal in minute 42, the market was closed for a short period before observed stakes increased during halftime and remained elevated in the second half, closely aligning with the prediction, albeit with some fluctuation. Despite this volatility, all observations clearly lie within the 99\% predictive interval, providing no evidence of suspicious behaviour in this example match.

\begin{figure}[htp!]
    \centering
    \includegraphics[width=0.65\linewidth]{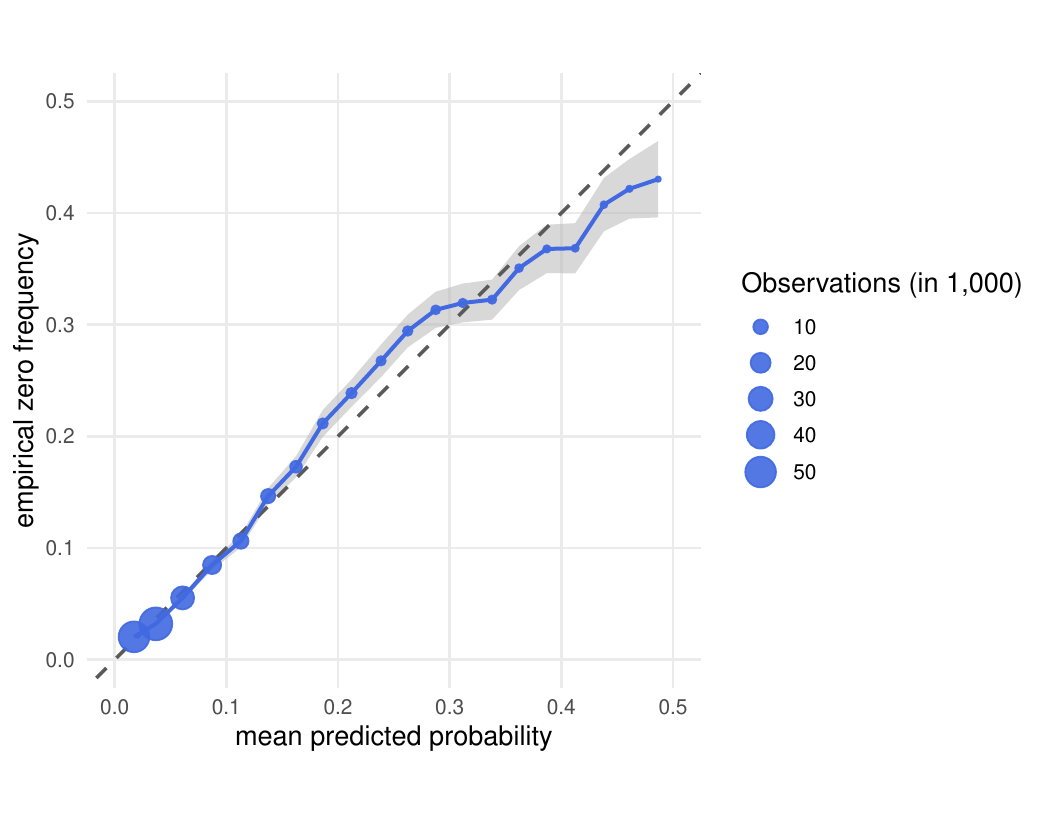}
    \caption{Calibration plot of theoretical and empirical probability to observe zero stakes despite an open market, including 95\%~confidence intervals. Point sizes indicate the number of underlying observations in the respective bin.}
    \label{fig:zero_calibration}
\end{figure}

In the second step, we consider the binary component used to predict the probability of observing zero stakes in an open market. Figure~\ref{fig:zero_calibration} shows a calibration plot comparing mean predicted probabilities (x-axis) with the empirical frequencies (y-axis), together with their 95\%~confidence intervals. Predicted probabilities are grouped into 2.5-percentage-point bins, with point sizes proportional to the number of observations per bin (for better readability, we exclude those 0.07\% of observations where the predicted probability of observing zero stakes exceeds 0.5). Most observations fall into bins where both predicted and empirical zero frequency are close to zero. For empirical frequencies between 0.2 and 0.3, the model slightly underestimates the probability of zero stakes, whereas it tends to slightly overestimate it for frequencies above 0.4. Since our primary objective is to identify suspiciously large observations, the zero-stake component is of minor importance, and its fit is sufficiently precise for our purpose. This is also reflected in the scoring rules: the Brier score is 0.074 (an improvement over a constant-rate benchmark of approximately 12.5\%) and the log loss is 0.260 (an improvement of approximately 16.4\%).

\section{Outlier analysis}
\label{sec:outlier}
To address the paper’s central questions -- namely, how to safeguard integrity in sports by detecting suspicious betting patterns that may be consistent with match-fixing -- we closely examine the pseudo-residuals $r_t$ introduced in Section~\ref{sec:checking}. For each match, observation, and outcome, these pseudo-residuals capture the discrepancy between the stakes predicted by our model and the stakes observed in the market. This is equivalent to flagging observations in which realised stakes exceed a given threshold -- effectively a quantile -- in the model's predictive distribution, thereby enabling bookmakers to react in-play. Our primary focus is on extreme positive deviations, as corresponding outliers indicate unusually large volumes placed on a specific outcome relative to what would be expected given the available information, which may therefore warrant further investigations into potentially irregular activity. For the following analysis, we consider all stakes that exceed the 99.8\%~quantile, yielding 195 observations (0.09\% of all observations in which the market was open; for the relation between the chosen threshold and the number of flagged observations, see Figure~\ref{fig:flaggs} in Appendix~\ref{sec:app_flaggs}). These 195 observations are associated with 175 distinct time series of individual teams in single matches (match-team-series, cf.\ Section~\ref{sec:framework}). The flagged series involve 174 matches; there is thus only a single match in which betting on both teams is flagged, and only 16 series feature two or more unusually large stakes (maximum: four).

To identify potentially recurring patterns, we compare the 195 flagged observations with all observations for which placing bets was possible (i.e.\ when the market was open). Since no confirmed cases of match-fixing are documented for the matches in our sample, this comparison serves to characterise the situations in which such large deviations occur, rather than to establish them as indicators of manipulation. Flagged observations disproportionately often involve bets on away teams (64.6\%) and on teams that are one goal behind (25.1\% compared with 18.7\% in the full dataset), while teams one goal ahead are underrepresented (4.1\% compared with 18.7\%). Unsurprisingly, this coincides with a lower mean implied win probability (23.6\% compared with 35.2\% overall); in 76.4\% of cases, the implied win probability is lower than at kick-off. The mean stake on the flagged outcome is 23.25, approximately 19 times the overall mean of 1.19. 

\begin{figure}[htp!]
    \centering
    \includegraphics[width=0.8\linewidth]{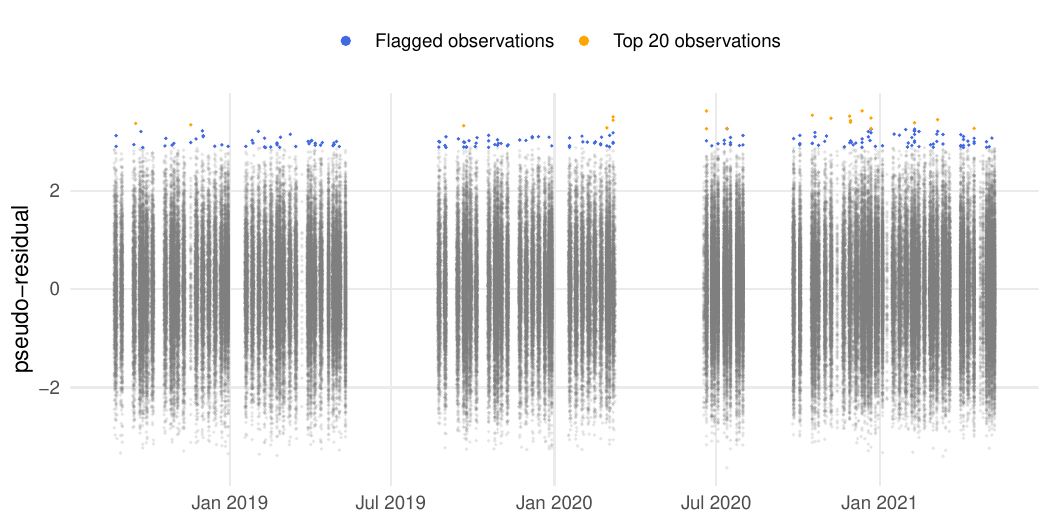}
    \caption{Pseudo-residuals over calendar time. Observations with stakes exceeding the 99.8\% and 99.94\%~quantile of their respective one-step-ahead predictive distributions are coloured.}
    \label{fig:res_calendar}
\end{figure}

\begin{figure}[htp!]
    \centering
    \includegraphics[width=0.6\linewidth]{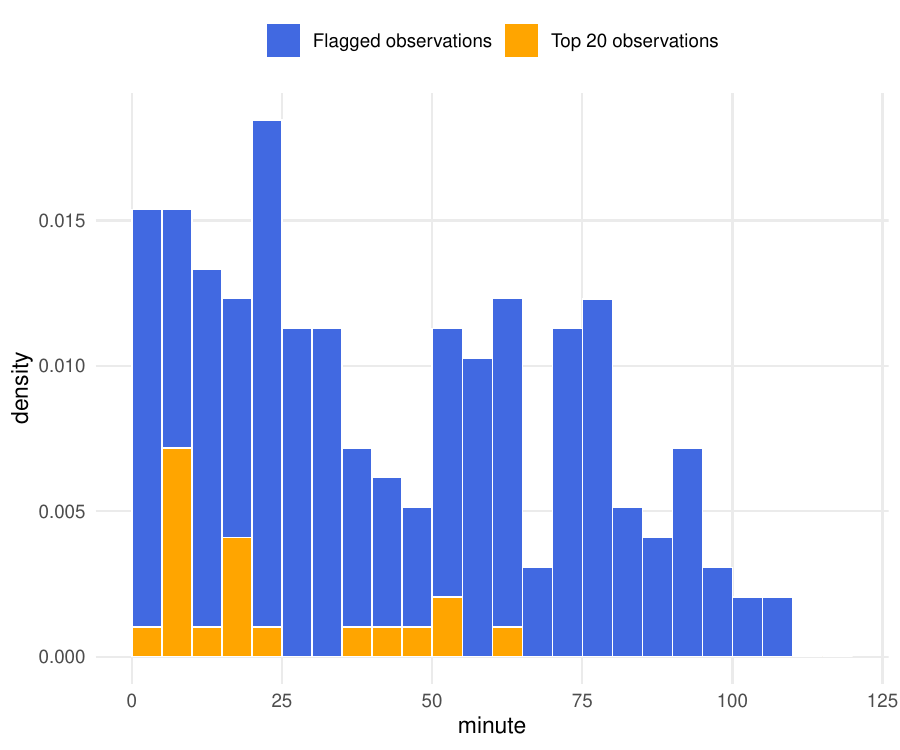}
    \caption{Distribution of flagged observations over the course of the match}
    \label{fig:res_minute}
\end{figure}

Figure~\ref{fig:res_calendar} plots pseudo-residuals against calendar date. Flagged observations occur across all three seasons in our sample and throughout all periods of the season. However, their frequency appears to increase slightly in the 2020/21 season (40\% of the flagged observations stem from this season). With respect to the minute of the match, Figure~\ref{fig:res_minute} shows that flagged stakes are more common in the first half (55.4\%) and up to the end of halftime (68.7\%). This exceeds the corresponding share of overall betting activity (39.3\% during the first half; 53.3\% until the end of halftime), as reflected in the average team stakes in Figure~\ref{fig:stakes_minute}. Relating flagged observations to the number of matches played, Juve Stabia (7 of 38 matches), Virtus Entella (13 of 76), and Perugia (11 of 74) are flagged most frequently. By contrast, Palermo and Hellas Verona, which stand out in terms of average stakes per minute (cf.\ Figure~\ref{fig:teamsavgstakes} in Appendix~\ref{sec:app_teamstakes}), are flagged only once and not at all, respectively.

\begin{figure}[htp!]
    \centering
    \includegraphics[width=0.7\linewidth]{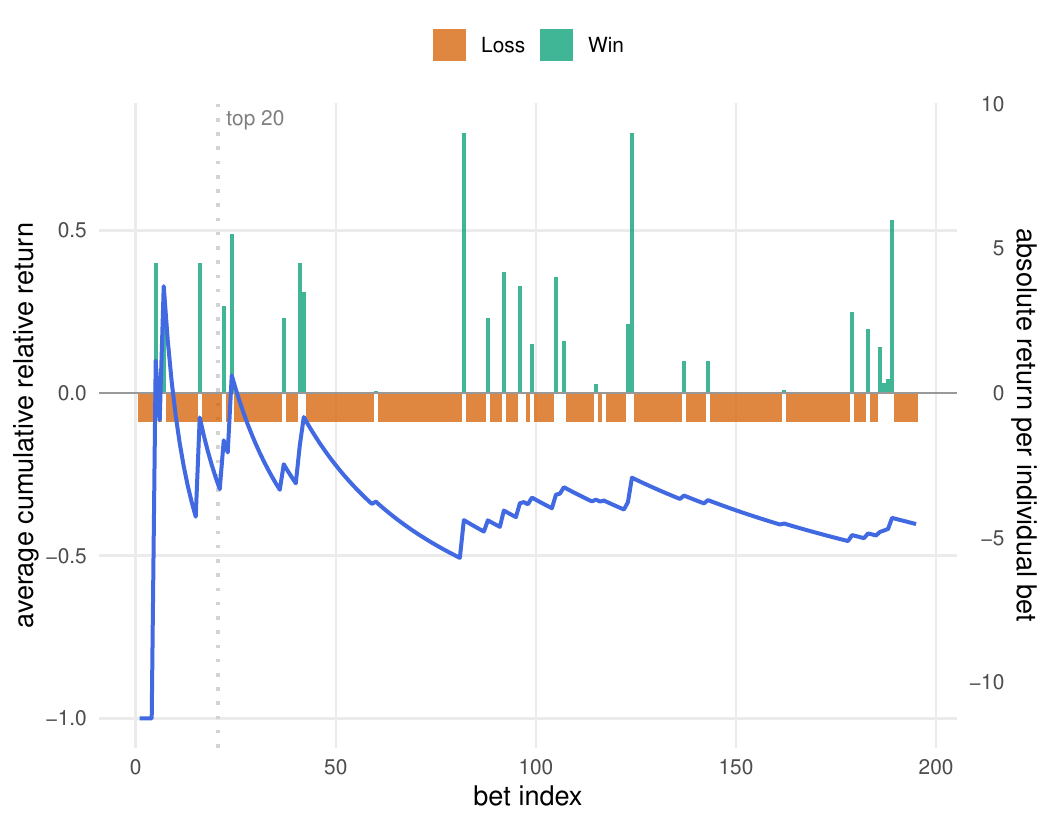}
    \caption{Absolute return per \euro{} wagered in individual bets (right y-axis) and average cumulative relative return on investment for the bets with the x~highest pseudo-residual values in the dataset, when placing the same stakes on each bet (blue line; left y-axis). For example, the average cumulative relative return on investment for the 50 bets with the highest pseudo-residual values is -22.2\%. The grey-dotted line indicates the top 20 flagged observations.}
    \label{fig:return}
\end{figure}

Turning to outcomes, among the 175 flagged series, 69 matches ended in a home win (39.4\%; 42.5\% in the full dataset), 56 in a draw (32.0\%; 31.3\%), and 50 in an away win (28.6\%; 26.2\%). Accordingly, the empirical distributions of flagged match-team-series and of all matches are very similar. Within three minutes after a flagged observation, the backed team scores in 2.6\% of cases, while the opposing team scores in 6.2\% (across all observations, a goal occurs within the subsequent three minutes in 3.1\% of the cases). At a five-minute horizon, the corresponding figures are 3.6\% (backed team scores) and 8.2\% (opponent scores), compared with 5.2\% across all observations. The backed team wins in only 30 of the 175 flagged match-team-series (17.1\%). Figure~\ref{fig:return} reports the absolute return per \euro{} wagered in individual bets (right y-axis) and the average relative accumulated return on investment when placing the same stake on the bets with the top-x pseudo-residual values (left y-axis). The high odds on some bets lead to a positive average cumulative return at a few points among the first 25 observations. Nevertheless, with only a few winning bets among all flagged observations, it is unsurprising that the average return on investment of -40\% is clearly negative and falls below the average bookmaker margin of approximately 8.7\% across all observations where betting was possible.

Finally, we examine the 20 observations with the highest pseudo-residual values (those exceeding the 99.94\%~quantile of the predictive distribution) in more detail. Table~\ref{tab:outliers_details} reports the corresponding match and date, the predicted and actual stakes, and key match characteristics. Seventeen out of 20 bets are on away teams, and all are placed before or during halftime at a score of either 0-0 or 1-0. Fifteen out of 20 observations occur between 07 March 2020 and 07 March 2021 (cf.\ Figure \ref{fig:res_calendar}). To check whether seasonal differences in betting activity explain this pattern, we repeat the analysis using season-specific rather than overall average team stakes in Equation~\ref{eq:covariates_mean}. However, the 20 identified matches remain the same. Moreover, all bets are on the underdog or very slight favourite: the implied win probability is below 25\% in 15 out of 20 cases. All realised amounts clearly exceed the 99\%~quantile of the predicted distribution by a wide margin (up to a factor of more than 160). Three out of the 20 bets are won, implying a return on investment of -26\% under constant wagering across matches. This is again considerably less than the average bookmaker's margin. Regarding match outcomes, nine out of the 20 matches with the most extreme observations ended in a draw.

\begin{table}[htp!]
    \centering
    \scalebox{0.6}{
    \begin{tabular}{ccc|cccc|cc|ccc}
         home & away &  &  & bet on & current & final & predicted stake & actual &  & implied win & bet  \\
         team & team & date & minute & team & score & score & (99\%~quantile) & stake & odds & probability (\%) & won \\
        \hline
        \textbf{AS Livorno}          & FC Crotone            & 17.09.2018 & 6 & home & 0-0 & 0-1 & 1.43 & 48.99 & 2.9 & 31.50 & no \\ 
        Spezia Calcio       & \textbf{Benevento Calcio}      & 18.11.2018 & 21 & away & 1-0 & 3-1 & 1.82 & 51.88 & 7.0 & 13.04 & no \\ 
        AS Livorno          & \textbf{Pordenone Calcio}      & 21.09.2019 & 54 & away & 1-0 & 2-1 & 0.54 & 12.95 & 13.0 & 7.13 & no \\ 
        AC Chievo Verona    & \textbf{AS Livorno}            & 29.02.2020 & 36 & away & 0-0 & 0-1 & 2.34 & 51.52 & 5.5 & 16.64 & yes \\ 
        Frosinone Calcio    & \textbf{US Cremonese}         & 07.03.2020 & 10 & away & 0-0 & 0-2 & 0.62  & 51.33  & 5.5 & 16.80 & yes \\
        FC Empoli           & \textbf{Trapani Calcio}       & 07.03.2020 & 44 & away & 0-0 & 1-1 & 1.26  & 64.93  & 5.5 & 16.41 & no \\
        US Salernitana 1919 & \textbf{Pisa SC}              & 20.06.2020 & 7  & away & 0-0 & 1-1 & 0.62  & 103.05 & 3.8 & 24.00 & no \\
        \textbf{AS Livorno}          & AS Cittadella         & 20.06.2020 & 62 & home & 0-0 & 0-2 & 5.21 & 102.8 & 4.5 & 20.35 & no \\ 
        Cosenza Calcio      & \textbf{Perugia Calcio}        & 13.07.2020 & 8 & away & 0-0 & 2-1 & 5.29 & 103.28 & 2.6 & 34.97 & no \\ 
        Virtus Entella      & \textbf{US Reggina 1914}      & 17.10.2020 & 53 & away & 0-0 & 1-1 & 1.15 & 103.10 & 3.8 & 23.49 & no \\
        US Cremonese        & \textbf{Vicenza Virtus}       & 07.11.2020 & 7  & away & 0-0 & 0-1 & 1.17 & 77.02  & 3.8 & 24.00 & yes \\
        AC Monza            & \textbf{US Reggina 1914}      & 28.11.2020 & 50 & away & 0-0 & 1-0 & 1.13  & 77.55  & 7.5 & 12.24 & no \\
        \textbf{Cosenza Calcio}      & US Salernitana 1919  & 29.11.2020 & 2  & home & 0-0 & 0-1 & 2.54 & 102.71 & 2.7 & 34.94 & no \\
        AC Reggiana         & \textbf{US Cremonese}          & 29.11.2020 & 19 & away & 1-0 & 1-1 & 1.55 & 52.72 & 5.0 & 18.48 & no \\ 
        US Cremonese        & \textbf{Ascoli Calcio 1898}   & 12.12.2020 & 16 & away & 0-0 & 3-3 & 0.33  & 51.41  & 5.0 & 18.18 & no \\
        FC Empoli           & \textbf{AC Reggiana}          & 22.12.2020 & 18 & away & 0-0 & 0-0 & 0.75  & 51.35  & 6.5 & 14.07 & no \\
        Cosenza Calcio      & \textbf{Venezia}               & 22.12.2020 & 6 & away & 0-0 & 0-0 & 3.75 & 77.21 & 2.7 & 34.15 & no \\ 
        US Lecce            & \textbf{Brescia Calcio}        & 09.02.2021 & 13 & away & 0-0 & 2-2 & 0.47 & 20.53 & 5.0 & 18.14 & no \\ 
        FC Empoli           & \textbf{AS Cittadella}        & 07.03.2021 & 9  & away & 0-0 & 1-1 & 0.38  & 20.23  & 4.3 & 21.18 & no \\
        AC Monza            & \textbf{US Cremonese}          & 17.04.2021 & 16 & away & 0-0 & 2-1 & 0.72 & 15.49 & 3.4 & 27.18 & no \\ 
    \end{tabular}}
    \caption{Matches, characteristics of the bet, and predicted as well as empirical stakes for the 20 observations with the highest pseudo-residual value. Teams backed are highlighted.}
    \label{tab:outliers_details}
\end{table}

\begin{figure}[htp!]
    \centering
    \includegraphics[width=0.8\linewidth]{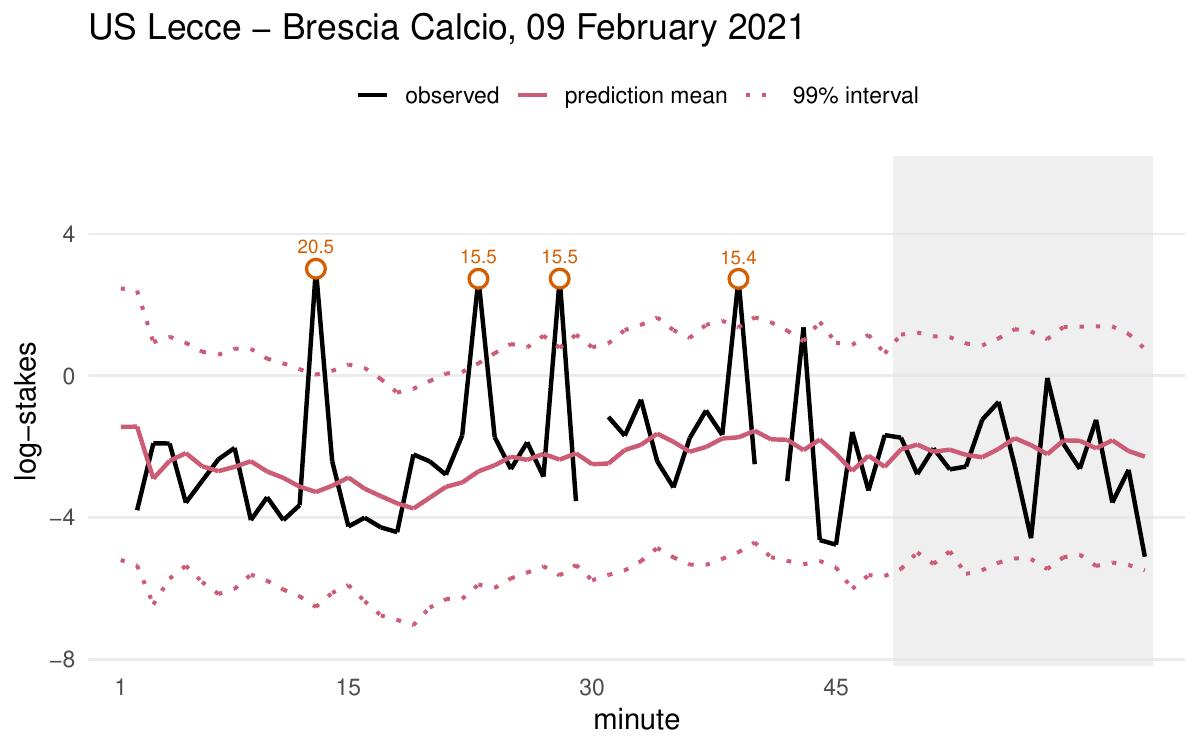}
    \caption{One-step-ahead predictions together with corresponding intervals covering the central 99\% of all predictions for log-stakes placed on Brescia Calcio in their away match against US Lecce on 09 February 2021. The figure shows only the period until the end of the halftime, as only 2\% of total stakes were placed during the second half, with numerous observations in which no stakes were placed. Unusual betting activity (stakes exceeding the 99.8\%~quantile of the predictive distribution) is highlighted, along with the actual betting stakes on the original scale. The grey-shaded area corresponds to halftime.}
    \label{fig:match_pred_outlier}
\end{figure}

The teams AS Livorno and US Cremonese each appear in three cases as the team backed. However, only one bet on each was successful. In addition, the match between US Lecce and Brescia Calcio on 09 February 2021 is of particular interest: four pseudo-residual values for bets on Brescia Calcio exceed our 99.8\%~quantile threshold, and a fifth is close. In this match, US Lecce (the home team) scored in the 48th and 56th minutes of the match (note that these differ from the values of \textit{minute} in our dataset, which also counts the time elapsed during half-time). The away team, Brescia Calcio, scored in the 75th minute of the match and equalised late in the injury time added to the second half. Our analysis of pseudo-residuals indicates extreme betting activity on Brescia Calcio in the 13th, 23rd, 28th, 39th, and 43rd minutes. However, less than 2\% of the stakes on this team were placed after US Lecce scored the opening goal at the very beginning of the second half. Figure~\ref{fig:match_pred_outlier} shows the predicted stakes, including 99\%~confidence intervals, together with the actual stakes until the end of the halftime. The outlying betting stakes are highlighted. Although Brescia Calcio eventually equalised, all bets on a win for this team were lost. 

\section{Conclusion}
\label{sec:conclusion}
We developed a state-space modelling framework for high-frequency betting stakes in the Italian Serie~B, using data from the 2018/19–2020/21 seasons provided by a leading European bookmaker. Incorporating covariates that capture team- and match-specific characteristics, motivated by existing literature and a descriptive analysis, the model provides predictions of expected betting volumes and their distributions for every minute of each match. We find that bettors tend to place higher stakes on pre-match favourites and on teams in the lead, whereas a red card is associated with lower average stakes on the affected team and higher stakes on the opponent. The underlying (latent) level of betting activity responds significantly to goals: surprising goals increase activity, while unsurprising goals decrease it. These predictions form the basis for comparing model-implied and empirical stakes through pseudo-residuals, thereby facilitating the identification of abnormal betting activity.

We classify an observation as potentially suspicious if the realised betting stakes exceed the 99.8\%~quantile of our model's predictive distribution. Among these flagged bets, no clear patterns could be identified. The relative frequency of flagged observations is somewhat higher before and up to the end of halftime, and 17 of the 20 largest pseudo-residuals concentrate between February 2020 and April 2021. While bets on away teams and on teams trailing by one goal, and therefore those with lower implied win probabilities, are overrepresented, the distribution of match outcomes among flagged bets closely matches that of the full sample. In particular, we do not find evidence that teams attracting unusually large stakes have a higher probability of scoring in the subsequent minutes; if anything, the opposite holds: in these situations, the opponent turned out to be more likely to score than in the full dataset. Consistent with this, the average return on investment from wagering the same amount on all flagged bets is clearly negative and below that of a typical benchmark strategy (although a few bets did yield high positive returns). Hence, while the Italian Serie~B has historically been affected by match-fixing and fraudulent betting activity, our analysis does not provide evidence that the identified anomalies reflect match-fixing in our sample. Nevertheless, our proof-of-concept approach provides a promising tool for monitoring betting activity and identifying potentially suspicious behaviour in real time, thereby demonstrating how advanced statistical modelling can support the early detection of irregular betting patterns and strengthen integrity assurance in in-play (live) sports betting markets.

There are several natural ways to extend our study. First, our data covers only three seasons of Italy’s Serie~B. Although the league has experienced known match-fixing cases in the past, no fixed matches are known to have occurred in our sample. We therefore lack ground-truth data against which to assess whether the flagged observations correspond to actual manipulation. If suitable data become available, future research could evaluate the approach across seasons or leagues with confirmed cases, including earlier seasons of Serie B or recent cases in the Turkish league, and measure its ability to accurately identify such cases. Second, our predictions are generated in-sample. An out-of-sample evaluation, i.e.\ training the model on earlier seasons and assessing it on a subsequent season, would provide a more realistic test of its performance and better reflect how a bookmaker might use the framework to monitor matches in real time. Moreover, our outlier-detection procedure assesses observations individually rather than across consecutive periods. It may therefore flag individual spikes in betting volume but fail to detect a sustained increase in stakes spread across multiple observations. Future work could develop methods that detect such temporal patterns or take into account the actions of individual bettors, if such data were available. Finally, given the relatively small share of stakes placed on draws and the potentially greater difficulty of fixing a match to produce that outcome, our analysis is limited to bets on home and away teams and excludes bets on draws. Extending the model to include draw stakes would likely require altering the covariates and modelling assumptions.

\bigskip

\noindent
\textbf{Data availability statement}: The data used for this research comprises processed and analysed datasets derived from raw data provided by a major European bookmaker. The raw data itself is proprietary and cannot be shared due to contractual confidentiality agreements. \\

\noindent
\textbf{Generative AI statement}: The code was written from scratch by the authors. ChatGPT and Claude were used to assist in improving and structuring the code. The authors carefully reviewed the code and take full responsibility for its final version. In addition, during the preparation of this work, the authors used ChatGPT-5 to assist in improving language and clarity. The authors subsequently reviewed and edited the content to meet the required standards and take full responsibility for the final content of this publication. \\

\noindent
\textbf{Funding information}: This work was supported by the German Research Foundation (Grant 431536450), which is gratefully acknowledged.

\newpage
\bibliographystyle{apalike} 
\bibliography{library}

\newpage
\label{sec:appendix}

\appendix
\begin{appendices}

\section{Stake volumes per minute}
\label{sec:app_stakeslog}

\renewcommand{\thetable}{A\arabic{table}}
\renewcommand{\thefigure}{A\arabic{figure}}
\setcounter{table}{0}
\setcounter{figure}{0}

\begin{figure}[h]
    \centering
    \includegraphics[width=0.8\linewidth]{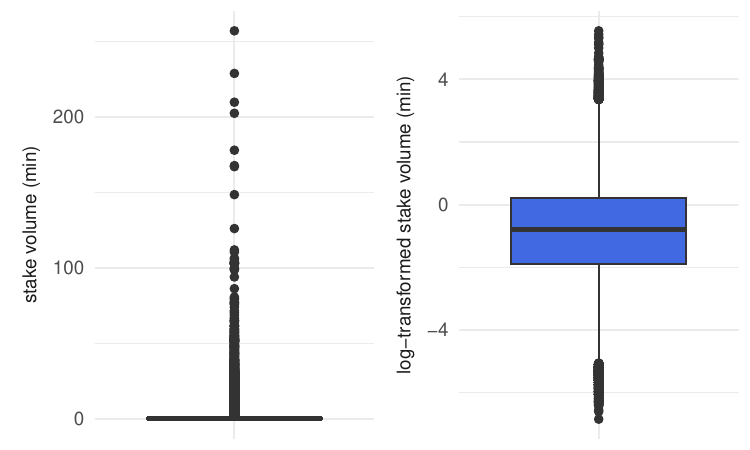}
    \caption{Distribution of stake volume per minute and its log-transformed version, shown as boxplots to compare the scale and skewness of the original and transformed data.}
    \label{fig:log_stakes}
\end{figure}

\section{Average team stakes}
\label{sec:app_teamstakes}

\renewcommand{\thetable}{B\arabic{table}}
\renewcommand{\thefigure}{B\arabic{figure}}
\setcounter{table}{0}
\setcounter{figure}{0}

\begin{figure}[htp!]
    \centering
    \includegraphics[width=0.7\linewidth]{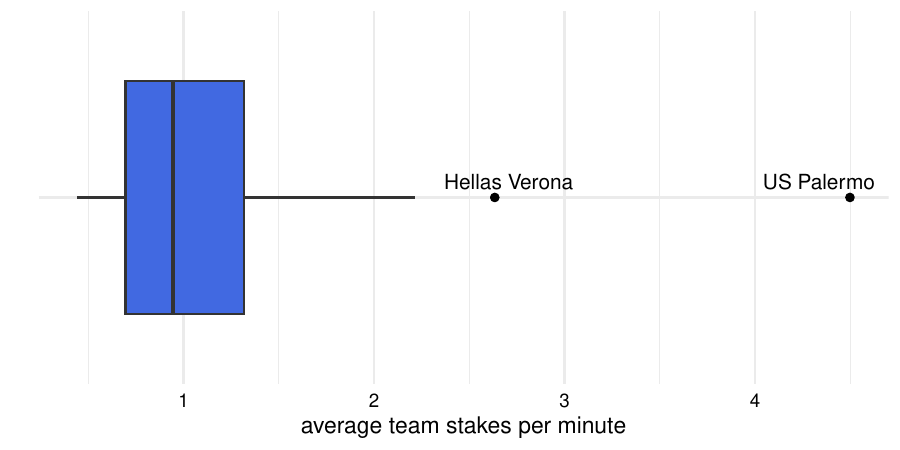}
    \caption{Boxplot of the average stakes placed per minute across all matches depending on the team.}
    \label{fig:teamsavgstakes}
\end{figure}

\newpage
\section{Total stakes per team}
\label{sec:app_totstakes}

\renewcommand{\thetable}{C\arabic{table}}
\renewcommand{\thefigure}{C\arabic{figure}}
\setcounter{table}{0}
\setcounter{figure}{0}

\begin{figure}[htp!]
    \centering
    \includegraphics[width=\linewidth]{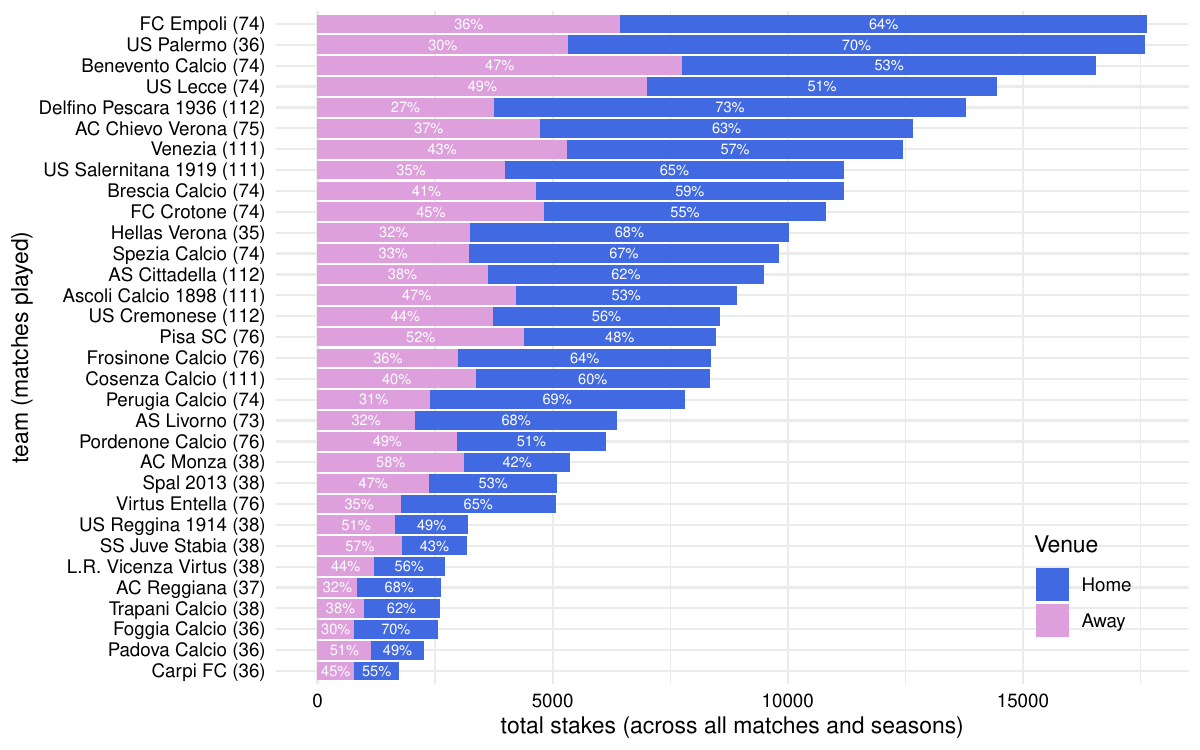}
    \caption{Total stakes placed on each team, split by whether the team played at home or away. Percentages give the home and away share of each team's total. The number of matches included differs across teams, since not all teams were in the league in all three seasons, and is given in the brackets after the team name.}
    \label{fig:teamstotstakes}
\end{figure}

\renewcommand{\thetable}{D\arabic{table}}
\renewcommand{\thefigure}{D\arabic{figure}}
\setcounter{table}{0}
\setcounter{figure}{0}

\newpage

\section{Stakes per scorediff}
\label{sec:app_stakesscorediff}

\begin{figure}[htp!]
    \centering
    \includegraphics[width=\linewidth]{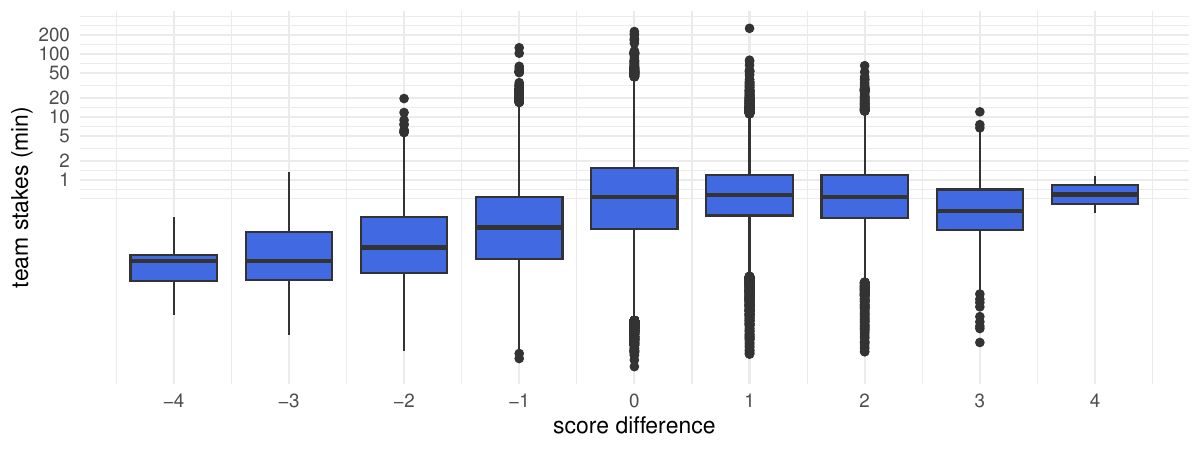}
    \caption{Boxplots of total stakes placed per minute by current score difference. Note that for visualisation purposes, we use a log scale on the y-axis.}
    \label{fig:volume_scorediff}
\end{figure}

\renewcommand{\thetable}{E\arabic{table}}
\renewcommand{\thefigure}{E\arabic{figure}}
\setcounter{table}{0}
\setcounter{figure}{0}

\section{Percentage of flagged observations depending on quantile threshold}
\label{sec:app_flaggs}

\begin{figure}[htp!]
    \centering
    \includegraphics[width=0.7\linewidth]{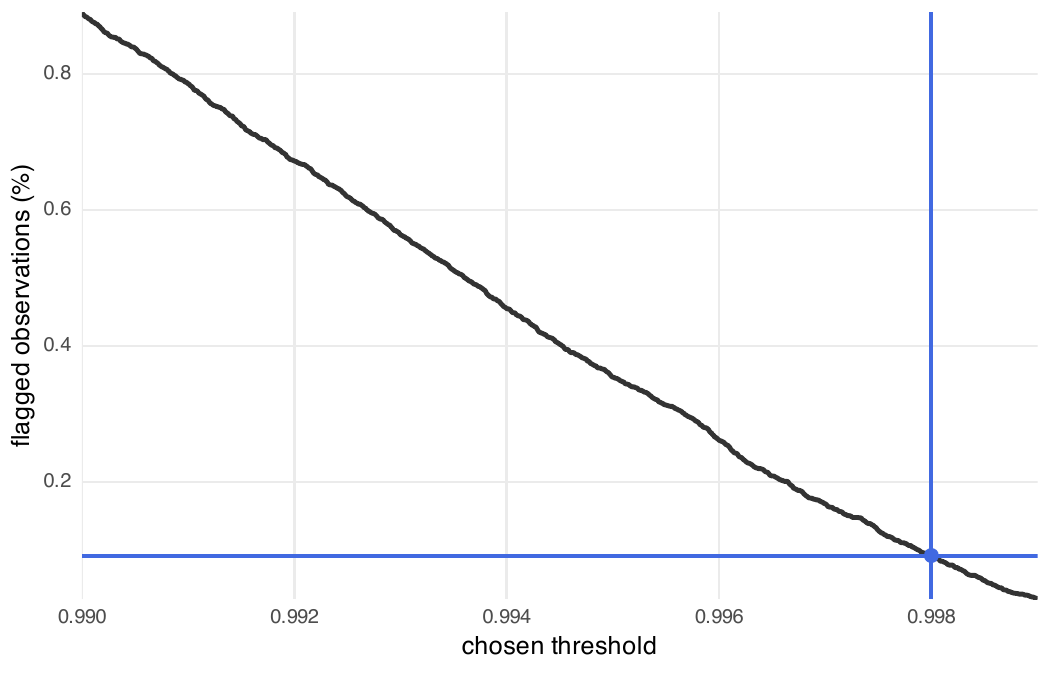}
    \caption{Percentage of flagged observations, depending on the chosen threshold of the model's predictive distribution of stakes placed.}
    \label{fig:flaggs}
\end{figure}

\end{appendices}

\end{spacing}
\end{document}